\documentclass[aps,prb,twocolumn,superscriptaddress,longbibliography,amsfonts,citeautoscript,nofootinbib]{revtex4-2}
\usepackage{amsmath, amssymb, graphicx, color, braket,bbold,multirow}
\usepackage[urlcolor=blue,colorlinks=true,citecolor=blue,linkcolor=blue,pdfstartview={FitH},bookmarks=false]{hyperref}

\begin{document}

\title{
Double-quantum-dot Andreev molecules:\\ Phase diagrams and critical evaluation of effective models
}

\author{Peter Zalom}
\email{zalomp@fzu.cz}
\affiliation{Institute of Physics, Czech Academy of Sciences, Na Slovance 2, CZ-18200 Praha 8, Czech Republic}

\author{Kacper Wrze\'{s}niewski}
\email{wrzesniewski@amu.edu.pl}
\affiliation{Institute of Spintronics and Quantum Information, Faculty of Physics and Astronomy, Adam Mickiewicz University in Pozna\'{n}, 61-614 Pozna{\'n}, Poland}

\author{Tom\'a\v{s} Novotn\'y}
\email{tomas.novotny@matfyz.cuni.cz}
\affiliation{Department of Condensed Matter Physics, Faculty of Mathematics and Physics, Charles University, Ke Karlovu 5, CZ-121 16 Praha 2, Czech Republic}

\author{Ireneusz Weymann}
\email{weymann@amu.edu.pl}
\affiliation{Institute of Spintronics and Quantum Information, Faculty of Physics and Astronomy, Adam Mickiewicz University in Pozna\'{n}, 61-614 Pozna{\'n}, Poland}

\date{\today}

\begin{abstract}

This work systematically investigates the phase diagram of a parallel double-quantum-dot Andreev molecule,
where the two quantum dots are coupled to a common superconducting lead.
Using the numerical renormalization group method,
we map out the evolution of the ground state across a wide parameter space of level detunings,  size of the superconducting gap, lead couplings, and inter-dot coupling strength. The intricate phase diagrams feature singlet, doublet, and a relatively uncommon triplet ground states, with the latter being a distinct signature of strong lead-mediated interactions between the quantum dots. We benchmark the applicability of simplified effective models, including the atomic limit and zero-bandwidth approximations, in capturing the complex behavior of this parallel configuration. Our analysis reveals severe limitations of these models, underscoring the necessity for maximal caution when extrapolating beyond their tested validity. In particular, all effective models except for the extended version of the zero-bandwidth approximation failed in reproducing the triplet ground state and made several false predictions. These findings provide crucial insights for interpreting experimental observations
and designing superconducting devices based on quantum-dot architectures.

 
\end{abstract}


\maketitle

Electrostatically defined quantum dots (QDs) in nanowires or carbon nanotubes, coupled with superconducting (SC) environments, offer a promising pathway for constructing quantum electronic circuits with diverse functionalities \cite{Buitelaar-2002, Graber-2004, Jarillo-Herrero-2006, Cleuziou-2006, vanDam-2006, Sand-Jespersen-2007, Pillet-2010, Alicea-2011, Rodero-2011, Delagrange-2015, Paaske-2015, Jellinggaard-2016, Wrzesniewski2017-kondo, Pankratova-2020, Klees-2020, Zalom-2021-reent, Bargerbos-2022, PitaVidal-2023,Zalom-2024}. These systems are characterized by the emergence of states within the SC gap which are referred 
to as either Yu-Shiba-Rusinov (YSR) or Andreev
bound states (ABS) \cite{Yu-1965, Rusinov-1969, Shiba-1973}, which are influenced by strong Coulomb interactions arising from their underlying electronic structure.
Consequently, a rich phenomenology of quantum phases is observed \cite{Yao2014Dec,Karrasch-2008,Meden-2019}.

While early efforts focused on single QD devices, advancements in fabrication have enabled the development of double-quantum-dot (DQD) systems in both serial and parallel configurations. Initially, parallel configurations predominantly appeared as central components of Cooper pair splitting (CPS) devices  \cite{Hofstetter-2009cps,DeFranceschi-2010,Herrmann-2010,Hoffstetter-2011cps,Schindele-2012cps,Fulop-2014cps,Fulop-2015cps,Tan-2015cps,deJong-2023cps}, correspondingly driving the bulk of the theoretical research into this particular direction \cite{Weymann-2014cps,Trocha-2015cps,Hussein-2016cps,Ranni-2021cps,Wrzesniewski-2017cps,Walldorf-2018cps,Wrzesniewski-2020cps, Bocian2018May,Brange-2021cps}. However, coupling two sites to a common reservoir following the Bardeen–Cooper–Schrieffer (BCS) theory has broader and more fundamental significance, as highlighted in recent proposal of quartet superconductivity at equilibrium \cite{Chirolli-2024}, in qubit engineering \cite{Ramsak-2006,Mishra-2021} or in minimal Kitaev chains \cite{Leijnse-2012,Tsintzis-2024}. A central concept is then the formation of Andreev molecular states, which have a non-local spatial nature, as recently studied experimentally in \cite{Su-2017,Kurtossy-2021,Kurtossy-2022,Matsuo-2023}.

Due to the strong electronic interactions various competing effects emerge in DQD devices at once. In SC serial configurations, Kondo screening and superconductivity shape simultaneously the outcome. Parallel DQD configurations have on top an inherent potential of an additional third effect, namely the lead-mediated interactions between the constituent QDs \cite{Eickhoff-2018}. In this regard, two opposing scenarios are particularly notable. 

\begin{figure}
	\includegraphics[width=0.95\columnwidth]{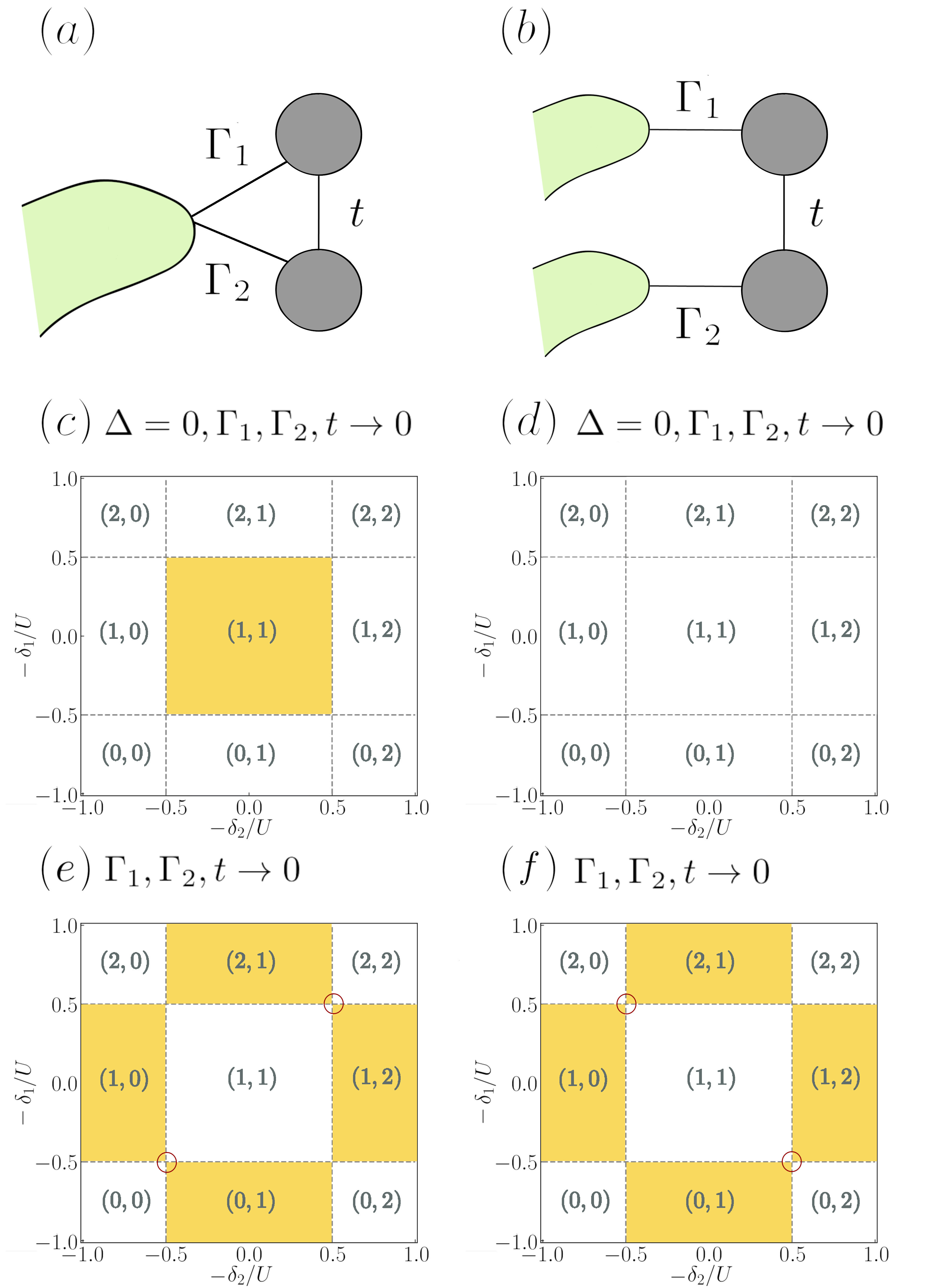}
	\caption{
		(a) Schematic of the considered system, consisting of two QDs
		(gray circles) coupled to a single BCS channel, with couplings $\Gamma_1$ and $\Gamma_2$
		and hopping between the dots denoted by $t$.
		Such scenario is achieved when the distance between QDs
		is much smaller than the superconducting coherence length $\zeta$.
		In the other limit, shown in (b), the DQD system can be well understood as a two channel model.
		(c) When $\Delta=0$ and $\Gamma_1, \Gamma_2, t \rightarrow 0$ the presence of only one screening channel
		leads to the formation of a doublet GS (yellow) in the middle of the phase diagram
		due to the underscreened Kondo effect. Changing the occupation of the dots by one electron
		leads then to the formation of singlet GS (white). The corresponding QD occupations $(n_1, n_2)$ are indicated.
		(d) For two screening channels, when $\Delta=0$, full screening of the DQD system is possible
		with the phase diagram consisting only of a singlet GS when the $\Gamma_1, \Gamma_2, t \rightarrow 0$.
		(e)-(f) For $\Delta$ being the dominant energy scale, no Kondo screening is possible and
		a checkerboard pattern of singlet and doublet GS emerges for one as well as two screening channels.
		Note that the location of the voided crossings (red circles) is distinctly different in both cases.
		Notice also that we show the phase diagrams as functions of {\it negative} detunings,
		which directly corresponds to experimental gate voltage sweeps.
		\label{fig:models}}
\end{figure}

If the spatial distance between the two QDs is negligible
compared to the SC coherence length $\zeta$ of their common lead,
one may assume that a single screening channel is available to both QDs.
Thus, a theoretical model of one BCS lead connected simultaneously to two QDs may be formulated, as schematically shown in Fig.~\ref{fig:models}(a). In the opposite limit, one may imagine the two QDs being physically far apart. Then, even if connected to a common lead, they will practically experience two independent screening channels already in the metallic case \cite{Eickhoff-2018}. Consequently, two BCS channels are required in the presence of superconducting correlations in the leads. They are each coupled to a separate QD, as shown in Fig.~\ref{fig:models}(b).

In the latter scenario, all lead-mediated interactions between QDs disappear and a relatively well-understood serial DQD scenario emerges effectively. Here, an extensive theoretical understanding already exists, which encompasses accurate non-perturbative results obtained via the Functional and Numerical Renormalization Group (NRG) or Quantum Monte Carlo (QMC) methods \cite{Zitko-2010,Zitko-2015,Zonda-2023,Karrasch-2011}%
\footnote{If the study is limited only to the subgap properties, the recently developed surrogate model \cite{Baran-2023} might also be applied.}. The accumulated systematic knowledge even allowed the development of numerically inexpensive effective descriptions via simplistic models, such as zero bandwidth (ZBW) \cite{Bergeret-2006,Grove-Rasmussen-2018}, atomic limit (AL) \cite{Oguri-2004-al,Tanaka-2007,Meng-2009} and its generalized version (GAL) \cite{Zonda-2023}. While in ZBW, the $\mathbf{k}$-dependency of the lead is dropped, describing effectively all screening effects via single orbital, AL can be viewed as a $\Delta \rightarrow \infty$ description. Adding proper parameter rescallings in AL leads then to the GAL theory.

In the one-SC-channel scenario, the current understanding is considerably less systematic \cite{Ramsak-2006,Minchul-2010,Yi-2018dqd}, with effective models often substituting the full descriptions, despite their extensions not being sufficiently tested. Using NRG \cite{Wilson-1975,Bulla-Rev-2008, Satori-1992, Sakai-1993,Zalom-2022,Zalom-2023} as our primary method, we therefore systematically solve the parallel DQD case across a wide parameter space, focusing particularly on regions with significant lead-mediated interactions between the QDs. 

The phase diagrams we find include singlet, doublet, and relatively uncommon triplet ground states (GSs) \cite{Karrasch-2011}, with the latter one being an important signature of the presence of significant lead-mediated interactions. Finally, we show that triplet phases can be reproduced by neither AL nor ZBW models, revealing thus their critical failure even in terms of qualitative predictions. This consequently underscores the necessity for increased caution and thorough cross-verification with either exact or well-defined approximate methods to avoid misleading conclusions when extrapolating AL or ZBW models to more complex QD systems.

The present paper is organized as follows.
First, the full model for the parallel DQD system with two Anderson orbitals is defined in Sec.~\ref{sec:anderson} with a corresponding tunneling self-energy derived in Sec.~\ref{sec:self_energy}.
Subsequently, the AL theory is formulated in Sec.~\ref{sec:al},
while the corresponding ZBW and its extension are defined in Sec.~\ref{sec:zbw}.
Employing NRG, we solve first for exact phase diagrams at zero inter-dot hopping (Sec.~\ref{sec:exact_t0})
and then discuss their development as the hopping is turned on (Sec.~\ref{sec:exact_tdep}).
The exact data provides then a substantial basis to disclose severe limitations of applying AL theory beyond the limitation of a large gap limit, as discussed in Sec.~\ref{sec:al_limits}. The same limitations are in principle observed for the application of ZBW, as shown in Sec.~\ref{sec:ZBW_fail}. Its extended version is then applied to model the triplet phase in Sec.~\ref{sec:eZBW}. The important features of the parallel DQD phase diagrams and their consequences for reliable modeling by simplified effective models are summarized in Sec.~\ref{sec:conslusions}.


\section{Theoretical framework \label{sec:theory} }

\subsection{SC-Anderson model for the parallel DQD scenario \label{sec:anderson}}

Our objective is to model a DQD screened by one spinfull SC channel, as shown schematically in Fig.~\ref{fig:models}$(a)$. We therefore consider an SC Anderson model with two QDs coupled to the same reservoir. The Hamiltonians for the upper ($j=1$) and lower QDs ($j=2)$ are then given as
\begin{eqnarray}
	H_{d,j}
	&=&
	\sum_{\sigma} 
	\left(
	\delta_{j}-\frac{U_j}{2}
	\right)
	d^{\dagger}_{j\sigma}
	d^{\vphantom{\dagger}}_{j\sigma}
	+
	U_j
	d^{\dagger}_{j\uparrow}
	d^{\vphantom{\dagger}}_{j\uparrow}
	d^{\dagger}_{j\downarrow}
	d^{\vphantom{\dagger}}_{j\downarrow},
\end{eqnarray}
with $d^{\vphantom{\dagger}}_{j\sigma}$ ($d^{\dagger}_{j\sigma}$) annihilating (creating)
electrons of spin $\sigma$ in the QD with label $j$. $U_j$ is the charging energy of the corresponding QD, $\varepsilon_j$ the energy level of dot $j$ and $\delta_j=\varepsilon_j+U_j/2$ represents its detuning from the half-filled single dot scenario, where $\delta_j=0$.

Two tunneling Hamiltonians, $H_{T,j}$ with $j \in \{1,2\}$,
allow for single-electron tunneling processes from the SC lead to
the $j$-th QD and vice-versa. They read
\begin{eqnarray}
	H_{T,j}
	&=&
	\sum_{\mathbf{k}\sigma}
	V_{j,\mathbf{k}}
	\left(
	c^{\dagger}_{\mathbf{k}\sigma}
	d^{\vphantom{\dagger}}_{j\sigma}
	+
	d^{\dagger}_{j\sigma}
	c^{\vphantom{\dagger}}_{\mathbf{k}\sigma}
	\right),
	\label{eq:lead_hop}
\end{eqnarray}
where $V_{j,\mathbf{k}}$ represent the corresponding spin-symmetric hybridization
terms for $j \in \{1,2\}$, which we assume for simplicity to be real.

Especially for experiments with two individual nanowires,
a direct inter-dot hopping cannot be avoided \cite{Kurtossy-2022}.
To model such a relevant situation, we define
\begin{eqnarray}
	H_t
	&=&
	t
	\sum_{\sigma}
	\left(
	d^{\dagger}_{1\sigma}
	d^{\vphantom{\dagger}}_{2\sigma}
	+
	d^{\dagger}_{2\sigma}
	d^{\vphantom{\dagger}}_{1\sigma}
	\right),
	\label{eq:interdot_hop}
\end{eqnarray}
with $t$ being the corresponding inter-dot hopping matrix elements. Here, we assume that the inter-dot charging energy
is negligible. Moreover, we also assume $U_1 = U_2 \equiv U$
and $\Gamma_1 = \Gamma_2 \equiv \Gamma$.
Finally, the SC lead is described via the ordinary BCS theory as
\begin{eqnarray}
	H_{\mathrm{SC}}
	&=&
	\sum_{\mathbf{k}\sigma} 
	\varepsilon_{\mathbf{k}}
	c^{\dagger}_{\mathbf{k}\sigma}
	c^{\vphantom{\dagger}}_{\mathbf{k}\sigma}
	+
	\Delta
	\sum_{\mathbf{k}}
	\left(
	c^{\dagger}_{\mathbf{k}\uparrow}
	c^{\dagger}_{\mathbf{k}\downarrow} 
	+
	h.c.
	\right),
	\label{eq:bcs_lead}
\end{eqnarray}
with $c^{\vphantom{\dagger}}_{\mathbf{k}\sigma}$ ($c^{\dagger}_{\mathbf{k}\sigma}$) being
the annihilation (creation) operator of the electron with spin $\sigma$, momentum $\mathbf{k}$, and energy $\varepsilon_k$ in the SC lead, while $\Delta$ stands for the gap parameter with a zero phase selected by using the BCS gauge invariance.

The total Hamiltonian for the DQD problem with one SC screening channel
is thus given as a sum of the previously-defined constituent parts:
\begin{align}
	H 
	= 
	\sum_{j \in \{1,2\}} \left( H_{d,j} 
	+ 
	H_{T,j}\right) 
	+ 
	H_t  + H_{\mathrm{SC}}.
	\label{eq:totalH} 
\end{align}

Although in this paper we examine the case of DQD coupled to a single BCS channel,
for comparison, let us qualitatively contrast the results against the two-channel case.
By comparing the one and two channel scenarios presented in Fig.~\ref{fig:models},
we can illustrate how lead-mediated interactions shape the phase diagrams
as $\delta_1$ and $\delta_2$ are varied.
Initially assuming $\Delta \rightarrow 0$, with each dot occupied by a single electron,
it becomes clear that a doublet GS forms in the one-channel case,
because one of the dots remains always unscreened
and the whole system exhibits underscreened Kondo effect.
In all other occupation regimes, a singlet GS appears,
as shown in Fig.~\ref{fig:models}(c).
For the odd DQD occupation this singlet is associated with
the development of the Kondo effect.
Conversely, the situation in the two-channel case
is markedly different, since both QDs can be simultaneously screened,
even when each is singly occupied. The phase diagram is then entirely composed of a singlet GS,
as illustrated in Fig.~\ref{fig:models}(d). Note, however,
that the nature of the GS can be completely different, depending on the DQD occupation.

Assuming now the SC effects to dominate, i.e. when the superconducting energy gap
is much larger than the Kondo temperature in all relevant regimes,
no screening takes place, but the ABSs form in the system.
Then, the single-electron occupation of both QDs results in a singlet GS
due to the disrupted and ineffective screening.
With both constituent QDs in a doublet GS, their overall parity becomes zero even for $t \neq 0$.
Changing the total occupation by one electron
results always in a parity changing transition to a doublet GS.
Consequently, a checkerboard pattern appears in both scenarios,
as shown in Figs.~\ref{fig:models}(e) and (f).
Different geometry of the two problems, however, leads to an avoided crossing
at different positions (marked by red circles), when $t$ is finite.
The evolution between the fully metallic and SC-dominated case
is thus expected to yield significantly different behavior,
providing thereby clear indicators of lead-mediated effects in the system.

\subsection{The tunneling self-energy \label{sec:self_energy}}

While the total Hamiltonian \eqref{eq:totalH} defines the present model fully, an alternative problem definition might be obtained by extracting the corresponding tunneling self-energy. 
To this end, let us follow the procedure in Ref.~\cite{Zalom-2024}, by taking advantage of the Nambu spinors
\begin{align}
	C^{\dagger}_{\mathbf{k}} 
	&= 
	\left(
	c^{\dagger}_{ \mathbf{k} \uparrow},
	c^{\vphantom{\dagger}}_{-\mathbf{k} \downarrow}
	\right),
	\\
	D_j^{\dagger} 
	&=
	\left(
	d^{\dagger}_{ j \uparrow},
	d^{\vphantom{\dagger}}_{j \downarrow}
	\right),
\end{align}
for $j=1,2$, and stack them into an infinite-dimensional vector
\begin{equation}
	\Psi^{\dagger}
	=
	(D_1^{\dagger},D_2^{\dagger},C^{\dagger}_{\mathbf{k}}),
\end{equation}
where $\Psi$ is its Hermitian conjugated counterpart in the form of a column vector. The spinors $C^{\dagger}_{\mathbf{k}}$ are understood to be repeated in $\Psi^{\dagger}$ for all possible quasi-momenta $\mathbf{k}$ of lead electrons.
In such a formalism, the non-interacting ($U=0$) Hamiltonian of the present model reads
\begin{equation}
	H_0
	=
	\Psi^{\dagger} 
	\mathbb{E}
	\Psi^{\vphantom{\dagger}},
\end{equation}
with
\begin{eqnarray}
	\mathbb{E}
	=
	\left(
	\begin{matrix}
		\mathbb{E}_{11}
		&
		\mathbb{E}_{12}
		&
		\mathbb{V}_{1 \mathbf{k}}
		\\
		\mathbb{E}_{21}
		&
		\mathbb{E}_{22}
		&
		\mathbb{V}_{2\mathbf{k}}
		\\
		\mathbb{V}_{1\mathbf{k}}
		&
		\mathbb{V}_{2\mathbf{k}}
		&
		\mathbb{E}^{\vphantom{\dagger}}_{\mathbf{k}}
	\end{matrix}
	\right).
	\nonumber
	\\
	\label{eq:mathbbE}
\end{eqnarray}
The blackboard bold typeface denotes $2 \times 2$ matrices:
\begin{align}
	\mathbb{E}_{\mathbf{k}}  
	& =
	-\Delta \sigma_x
	+ \varepsilon_{\mathbf{k}} \sigma_z,
	\\
	\mathbb{E}_{jj}  
	& =
	\left(
	\delta_j-\frac{U}{2}
	\right)\sigma_z,
	\\
	\mathbb{E}_{12}  
	&=
	\mathbb{E}_{21}  
	=
	t\sigma_z,
	\\
	\mathbb{V}_{j\mathbf{k}}  
	&=
	V_{j,\mathbf{k}}
	\sigma_z,
\end{align}
with $\sigma_x$ and $\sigma_z$ being the corresponding Pauli matrices. We then employ the partitioning scheme described in Appendix A of Ref.~\cite{Novotny-2005} to obtain the local retarded Green's function denoted as $\mathbb{G}_0(\omega^+)$, where $\omega^+\equiv \omega+i\eta$ with $\eta$ being infinitesimally small. It is a $4 \times 4$ matrix spanned by double-spinor $(D_1,D_2)$, which reads
\begin{eqnarray}
	\mathbb{G}_0(\omega^+)
	\hspace{-1.5mm}
	&=&
	\hspace{-1.5mm}
	\left[
	\omega^+ \mathbb{1}
	\hspace{-0.5mm}
	- 
	\hspace{-0.5mm}
	\left(
	\hspace{-1.0mm}
	\begin{array}{c c}
		\mathbb{E}_{11} & \mathbb{E}_{12}
		\\
		\mathbb{E}_{21} & \mathbb{E}_{22}
	\end{array}
	\hspace{-1.0mm}
	\right)
	\hspace{-0.5mm}
	- 
	\hspace{-0.5mm}
	\left(
	\hspace{-1.0mm}
	\begin{array}{c c}
		\mathbb{\Sigma}_{11}(\omega^+) & \mathbb{\Sigma}_{12}(\omega^+)
		\\
		\mathbb{\Sigma}_{21}(\omega^+) & \mathbb{\Sigma}_{22}(\omega^+)
	\end{array}
	\hspace{-1.0mm}
	\right)
	\right]^{-1},
	\nonumber
	\\
\end{eqnarray}
with the $2 \times 2$ blocks of self-energy given by
\begin{eqnarray}
	\mathbb{\Sigma}_{ij}(\omega^+)
	&=&
	\sum_{ \mathbf{k}}
	\mathbb{V}_{i,\mathbf{k}}
	\left(
	\omega^+ \, \mathbb{1}
	- 
	\mathbb{E}_{\mathbf{k}}
	\right)^{-1} 
	\mathbb{V}_{j,\mathbf{k}}.
	\label{eq:Sigma}
\end{eqnarray}
Resolving the matrix inversion of \eqref{eq:Sigma} leads to
\begin{align}
	\mathbb{\Sigma}_{ij}
	(\omega^+)
	=
	\sum_{\mathbf{k}}
	\frac{ 
		V_{i,\mathbf{k}} V_{j,\mathbf{k}}
		\left( \omega \mathbb{1} - \Delta \sigma_x + \varepsilon_{\mathbf{k}} \sigma_z \right)
	}
	{
		\omega^2 - \Delta^2 -\varepsilon_{\mathbf{k}}^2 + i\eta \, \mathrm{sgn}(\omega)
	}.
	\label{eq:Sigma_k}
\end{align}
Assuming now a constant tunneling density of states within the band of width $2D$
gives the final expression in the more convenient energy representation as
\begin{align}
	\mathbb{\Sigma}_{ij}
	(\omega^+)
	&=
	\frac{\Gamma_{ij}}{\pi}
	\int_{-D}^{D}
	\frac{ \omega \mathbb{1} - \Delta \sigma_x }
	{
		\omega^2 - \Delta^2 -\varepsilon^2 + i\eta \, \mathrm{sgn}(\omega)
	}
	d\varepsilon
	\nonumber
	\\
	&=
	\Gamma_{ij}
	\left( \omega \mathbb{1} - \Delta \sigma_x \right)
	F(\omega^+),
	\label{eq:Sigma_resolved}
\end{align}
where $\Gamma_{ij} = \pi \sum_{\mathbf{k}} V_{i,\mathbf{k}} V_{j,\mathbf{k}} \delta(\omega - \varepsilon_{\mathbf{k}})$ and $F(\omega^+)$ is a universal function given, for example, in Ref.~\cite{Zalom-2024}. To simplify the notation, the diagonal coupling strengths are further denoted as $\Gamma_{11} \equiv \Gamma_1$ and $\Gamma_{22} \equiv \Gamma_2$. In the end, each self-energy component $\mathbb{\Sigma}_{ij}$ possesses the same $2 \times 2$ Nambu matrix structure with only the corresponding coupling strength being different.

In real experiments the coupling strengths are dictated by geometrical aspects encoded in the $\mathbf{k}$-dependence of the hybridization terms. Consequently, for QDs separated by a distance much larger than the BCS coherence length, the off-diagonal couplings tend towards zero and the problem becomes then effectively of two-SC-channel nature, as depicted in Fig.~\ref{fig:models}(b). Once the QDs approach each other the sum $\sum_{\mathbf{k}} V_{i,\mathbf{k}} V_{j,\mathbf{k}} \delta(\omega - \varepsilon_{\mathbf{k}})$ becomes sizable. However, it still obeys Cauchy-Schwartz inequality. Thus, the off-diagonal coupling strengths satisfy $\Gamma_{12}=\Gamma_{21} \leq \sqrt{\Gamma_1 \Gamma_2}$. It is thus natural to parametrize the off-diagonal coupling strengths as $\Gamma_{12}=\Gamma_{21} = \nu \sqrt{\Gamma_1 \Gamma_2}$, with $\nu \in < 0,1 >$.

Clearly, lead-mediated correlations are proportional to the size of the off-diagonal coupling strengths. Concentrating especially on such effects in the paper, we therefore assume coupling symmetric scenario of $\Gamma_1=\Gamma_2$ with the largest possible off-diagonal couplings by selecting $\nu=1$. In more general scenarios with $\nu \neq 1$, when the underlying microscopic Hamiltonian is not known or available, the newly developed NRG diagonalization technique of Ref.~\cite{Zalom-2023} can be applied. This way, the transition from the present one-SC-channel model to the two-SC-channel one can be smoothly modeled.

\subsection{Atomic limit models \label{sec:al}}

A popular option for simplistic analysis of SC Anderson impurity models is represented
by the AL model or its most recent generalization GAL \cite{Zonda-2023}.
Their emergence as effective models is, however, not validated directly.
Instead, for single and serial DQDs,
its predictions have been verified to qualitatively agree with those obtained by either NRG or QMC \cite{Siano-2004,Meng-2009,Luitz-2010,Zitko-2015,Delagrange-2015,Grove-Rasmussen-2018,Saldana-2018,Zonda-2023}. In analytic terms, one takes the limit of infinite bandwidth followed by $\Delta \rightarrow \infty$ in the full self-energy $\Sigma$  \cite{Oguri-2004-al,Tanaka-2007, Meng-2009}. Thus, we obtain
\begin{align}
	\mathbb{\Sigma}^{\mathrm{AL}}_{ij}
	=
	\nu
	\sqrt{\Gamma^{\mathrm{AL}}_{i} \Gamma^{\mathrm{AL}}_{j} }
	\sigma_x,
	\label{eq:al_self_energy}
\end{align}
where using the superscript $\mathrm{AL}$ stresses that in practical applications the hybridizations $\Gamma_j^{\mathrm{AL}}$ are freely adjusted
so that either the exact NRG results \cite{Grove-Rasmussen-2018}
or the experimental data is reproduced.

Consequently, the AL model contains the diagonal blocks of environment's self-energy
describing direct Andreev reflections (DAR), which are proportional to
$\Gamma^{\mathrm{AL}}_j \equiv \Gamma_{\mathrm{DAR},j}^{\mathrm{AL}}$, and off-diagonal ones fixed to $\sqrt{\Gamma^{\mathrm{AL}}_1\Gamma^{\mathrm{AL}}_2} \equiv \Gamma_{\mathrm{CAR}}^{\mathrm{AL}}$, which in turn describe crossed Andreev reflections (CAR).
Once again, in practical applications, especially in the realm of CPS devices,
$\Gamma_{\mathrm{DAR},j}^{\mathrm{AL}}$ and $\Gamma_{\mathrm{CAR}}^{\mathrm{AL}}$ might be treated independently,
which models a scenario where the diagonal and off-diagonal blocks of the exact self-energy
of the environment can be considered as independent.

Due to the simple analytic form of the self-energy in the AL approximation,
it is straightforward to translate it into the corresponding Hamiltonian terms.
Given the two QDs remain described by $H_{d,1} + H_{d,2} + H_t$,
the AL model is given as a total Hamiltonian
\begin{align}
	H^{\mathrm{AL}}
	=
	H_{d,1} + H_{d,2}
	+
	H_t
	+ 
	H_{\mathrm{DAR}}
	+ 
	H_{\mathrm{CAR}},
	\label{eq:AL_dqd}
\end{align}
with
\begin{align}
	H_{\mathrm{DAR}}
	=
	&-\Gamma^{\mathrm{AL}}_1
	\left(
	d^{\dagger}_{1\uparrow} 
	d^{\dagger}_{1\downarrow} 
	+
	d^{\vphantom{\dagger}}_{1\downarrow} 
	d^{\vphantom{\dagger}}_{1\uparrow} 
	\right)
	\nonumber
	\\
	&
	-\Gamma^{\mathrm{AL}}_2
	\left(
	d^{\dagger}_{2\uparrow} 
	d^{\dagger}_{2\downarrow} 
	+
	d^{\vphantom{\dagger}}_{2\downarrow} 
	d^{\vphantom{\dagger}}_{2\uparrow} 
	\right),
\\
	H_{\mathrm{CAR}}
	=
	&-\nu\sqrt{\Gamma^{\mathrm{AL}}_1\Gamma^{\mathrm{AL}}_2}
	\left(
	d^{\dagger}_{1\uparrow} 
	d^{\dagger}_{2\downarrow} 
	+
	d^{\vphantom{\dagger}}_{1\downarrow} 
	d^{\vphantom{\dagger}}_{2\uparrow} 
	\right)
	\nonumber
	\\
	&
	-\nu\sqrt{\Gamma^{\mathrm{AL}}_1\Gamma^{\mathrm{AL}}_2}
	\left(
	d^{\dagger}_{2\uparrow} 
	d^{\dagger}_{1\downarrow} 
	+
	d^{\vphantom{\dagger}}_{2\downarrow} 
	d^{\vphantom{\dagger}}_{1\uparrow} 
	\right).
\end{align}
The diagonal blocks of the static self-energy contribute thus to $H_{\mathrm{DAR}}$,
while the off-diagonal blocks to $H_{\mathrm{CAR}}$. The AL theory approaches then $\Gamma^{\mathrm{AL}}_j$
as effective parameters, but does not provide a prediction of their values.
This shortcoming was recently mitigated by the GAL theory \cite{Zonda-2023},
which provided explicit prescriptions of the parameter rescalings in the case of single and serial DQD systems yielding thus good quantitative correspondence to the exact results.
We stress, however, that it shares with AL the same form of the effective Hamiltonian.
Consequently, if certain behavior is not present in the full parametric space of the AL Hamiltonian,
rescaling of the parameters cannot help in realizing it.
Summing up, the AL and GAL effective models are exact in the limit
of $\Delta \to \infty$. However,
since the gap is typically of the order of or smaller than the Coulomb correlations,
such limit is never reached in realistic experiments,
therefore these models should be used with a special care.

\subsection{Zero bandwidth approximations \label{sec:zbw}}

The zero bandwidth approximation provides another effective description
for numerically inexpensive analysis of Anderson models with SC leads.
However, we highlight that it has not been established on firm grounds
of either renormalization group (RG) or perturbation techniques,
nor does it follow from performing certain limits as in the case of AL theory.
Instead, its predictions have only been validated against the exact results
for single and serial DQDs attached to SC leads \cite{Grove-Rasmussen-2018}.
Yet, its simplicity made ZBW an increasingly popular tool even
in situations beyond the originally tested realm of applicability \cite{vonOppen-2021}.
Outside the original scope, however, ZBW results should be treated with caution, as mere extrapolations.

We therefore rigorously explore the potential of ZBW approximation
to model the present parallel DQD setup given by the full Hamiltonian \eqref{eq:totalH}.
To this end, we first lay down its ZBW approximation.
Following Ref.~\cite{Grove-Rasmussen-2018}, we drop the $\mathbf{k}$-dependence
in the BCS and tunneling Hamiltonians, leaving $H_{d,j}$ and $H_t$ unaltered.
We label the orbital representing the lead as $0$ and define its corresponding electron annihilation (creation) operator
as $c^{\vphantom{\dagger}}_{0\sigma}$ ($c^{\dagger}_{0\sigma}$). The total Hamiltonian in the ZBW approximation becomes then
\begin{align}
	H^{\mathrm{ZBW}}
	=
	\sum_{j \in \{1,2\}} 
	\left( 
	H_{d,j} + H_{T,j}^{\mathrm{ZBW}}
	\right)
	+
	H^{\mathrm{ZBW}}_{\mathrm{SC},0}
	+
	H_t,
	\label{eq:zbwH}
\end{align}
with
\begin{eqnarray}
H_{\mathrm{SC},0}^{\mathrm{ZBW}}
&=&
\Delta 
c^{\dagger}_{0\uparrow}
c^{\dagger}_{0\downarrow} 
+
h.c.,
\\
H^{\mathrm{ZBW}}_{j,\mathrm{T}}
\label{eq:BCS_zbw}
&=&
V_{j,0}^{\mathrm{ZBW}}
\sum_{\sigma}
\left(
c^{\dagger}_{0\sigma}
d^{\vphantom{\dagger}}_{j\sigma}
+
d^{\dagger}_{j\sigma}
c^{\vphantom{\dagger}}_{0\sigma}
\right).
\label{eq:tun_zbw}
\end{eqnarray} 
Notably, within the ZBW approximation the gap parameter $\Delta$ remains unaltered, however, $\mathbf{k}$-independent tunnel matrix elements, denoted here as $V_{j,0}^{\mathrm{ZBW}}$ for $j = 1,2$, are introduced. These have no direct relation to the bare parameters of the full Hamiltonian and it is understood that their values have an effective character. The ZBW theory, however, does not provide any analytic or numeric prescription for their values. Instead, $V^{\mathrm{ZBW}}_{j,0}$ is selected as a best possible fit to either precise theoretical results or experimental observations that the ZBW theory aims to replicate \cite{Grove-Rasmussen-2018}. We also emphasize that ZBW introduces
the corresponding coupling strengths as $\Gamma_{j,0}^{\mathrm{ZBW}}=\pi \left(V_{j,0}^{\mathrm{ZBW}}\right)^2/2$.
 
In accord with Ref.~\cite{Pavesic-2024}, a minimal extension of the ZBW approach to the present problem
is naturally achieved by adding an electronic orbital labeled $1$,
whose electron creation operator is denoted as $c^{\dagger}_{1\sigma}$.
The added orbital follows the Hamiltonian
\begin{align}
H_{\mathrm{SC},1}^{\mathrm{ZBW}}
=
\Delta 
c^{\dagger}_{1\uparrow}
c^{\dagger}_{1\downarrow} 
+
h.c.,
\end{align}
and is coupled directly only to the orbital of $c^{\dagger}_{0\sigma}$ electrons via the hopping Hamiltonian
\begin{eqnarray}
	H^{\mathrm{ZBW}}_{T,1}
	&=&
	V_{1}^{\mathrm{ZBW}}
	\sum_{\sigma}
	\left(
	c^{\dagger}_{0\sigma}
	c^{\vphantom{\dagger}}_{1\sigma}
	+
	c^{\dagger}_{1\sigma}
	c^{\vphantom{\dagger}}_{0\sigma}
	\right).
	\label{eq:tun_zbw_extend}
\end{eqnarray} 
The total Hamiltonian in this case, here referred to as the extended ZBW (eZBW), reads therefore
\begin{align}
	H^{\mathrm{eZBW}}
	=
	H^{\mathrm{ZBW}}
	+
	H^{\mathrm{ZBW}}_{\mathrm{T},1}
	+
	H^{\mathrm{ZBW}}_{\mathrm{SC},1}.
	\label{eq:ezbwH}
\end{align}

The ZBW approximation of the full Hamiltonian \eqref{eq:totalH} requires solving
a spinfull three-site model with a Hilbert space of $64$ states, while its minimal extension, eZBW, involves four Anderson orbitals with a Hilbert space of $256$ states.
Both therefore need significantly less computing resources compared to NRG or QMC calculations.
However, the lack of a prescription for the effective values in ZBW or eZBW requires
multiple runs to scan their parametric spaces, such that correct
effective values can be found to reproduce the exact results.

\begin{figure*}
	\includegraphics[width=1.95\columnwidth]{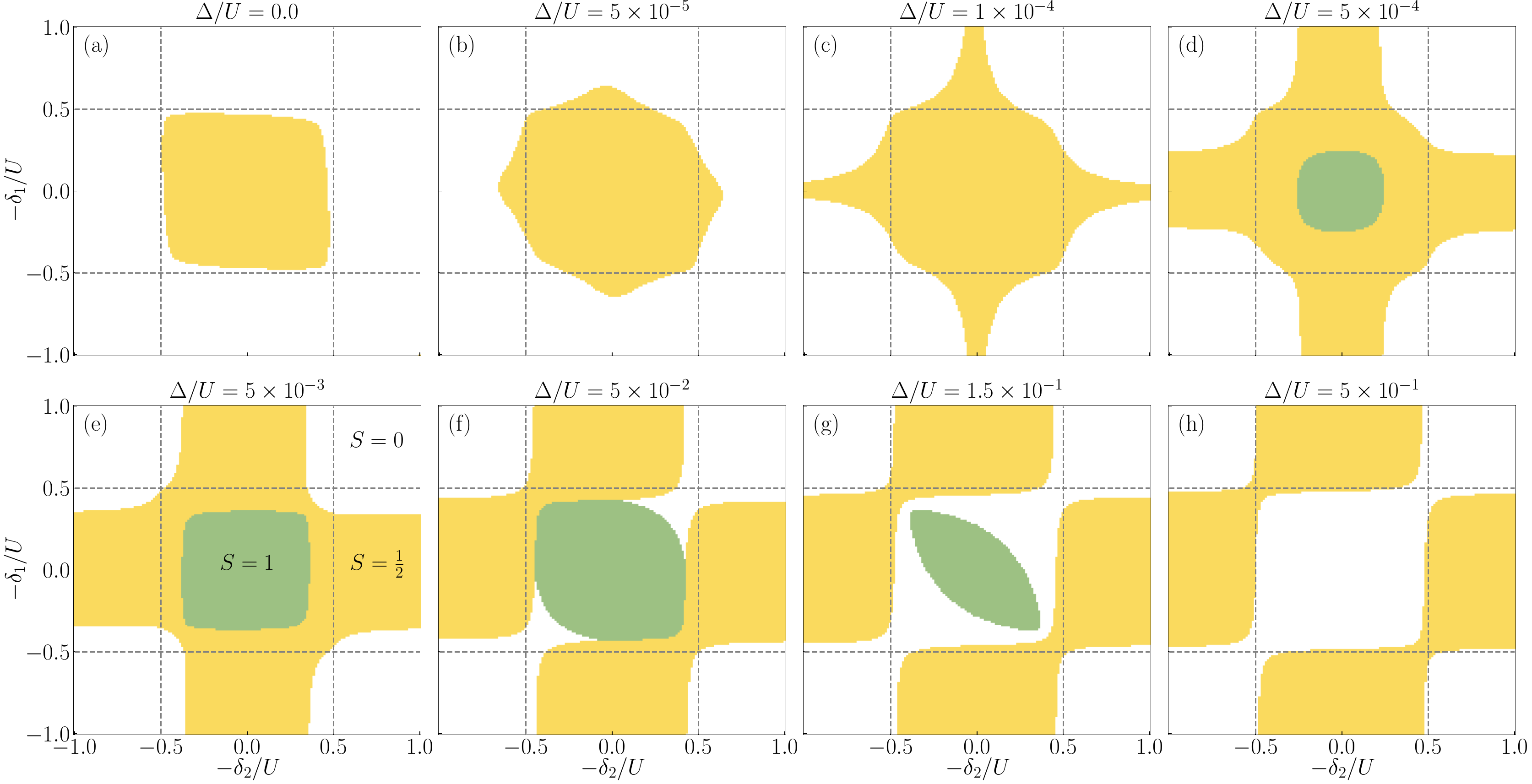}
	\caption{
		The NRG results for the phase diagrams for the DQD system of two identical QDs ($U_1=U_2\equiv U$), as functions of the detuning
		parameters $-\delta_1/U$ and $-\delta_2/U$ for different values of the superconducting energy gap $\Delta/U$,
		as indicated. Coupling-symmetric scenario $\Gamma_1/U=\Gamma_2/U\equiv \Gamma/U=0.05$ maximizing lead-mediated correlations ($\nu=1$) is selected.
		The white, yellow, and green regions correspond to singlet ($S=0$), doublet ($S=1/2$), and triplet ($S=1$) GS, respectively. (a) In the absence of superconductivity ($\Delta=0$), non-zero cross-correlations
		mediated by the common lead result in a doublet GS island at the center of the phase diagram.
		(b)-(c) 
		As $\Delta/U$ increases from $5\times10^{-5}$ to $10^{-4}$,
		the doublet GS island elongates into a cross-shaped region.
		(d)-(e) 
		For $\Delta/U=5\times10^{-4}$, a triplet GS emerges in the center of the phase diagram.
		(f)-(g) 
		The cross-shaped doublet region splits into two separate
		stripes along the main diagonal $\delta_1=\delta_2$,
		and the triplet region begins to shrink.
		(h) As $\Delta/U$ increases further to $0.5$, the triplet GS disappears,
		and a striped pattern of singlet and doublet phases
		appears in the direction of the main diagonal $\delta_1=\delta_2$.
		All calculations have been performed via NRG
	    with discretization parameter $\Lambda=2$
	    exploiting full spin symmetry with 
	    at least $1000$ multiplets kept during calculations,
		assuming $U=0.2D$ and the band halfwidth $D \equiv 1$ used as energy unit.
		\label{fig:nrg_UG_20}
}
\end{figure*}

\section{NRG results}

In this section we present and discuss the accurate NRG results
for the phase diagrams of the considered double quantum dot setup
as functions of orbital level positions. 
To perform the calculations, we have employed
standard NRG \cite{Wilson-1975, Krishna-1980a, Krishna-1980b, Bulla-1994, Bulla-Rev-2008, Satori-1992, Sakai-1993},
making use of the open-access
Flexible DM-NRG code \cite{fnrg}. We have assumed the discretization
parameter $\Lambda = 2$ and kept at least $1000$ multiplets at each iteration,
exploiting the full spin $SU(2)$ symmetry of the model.
These settings are applied in all NRG calculations presented in this paper. 

\subsection{Results in the absence of inter-dot hopping \label{sec:exact_t0}}

\begin{figure*}
	\includegraphics[width=1.0\textwidth]{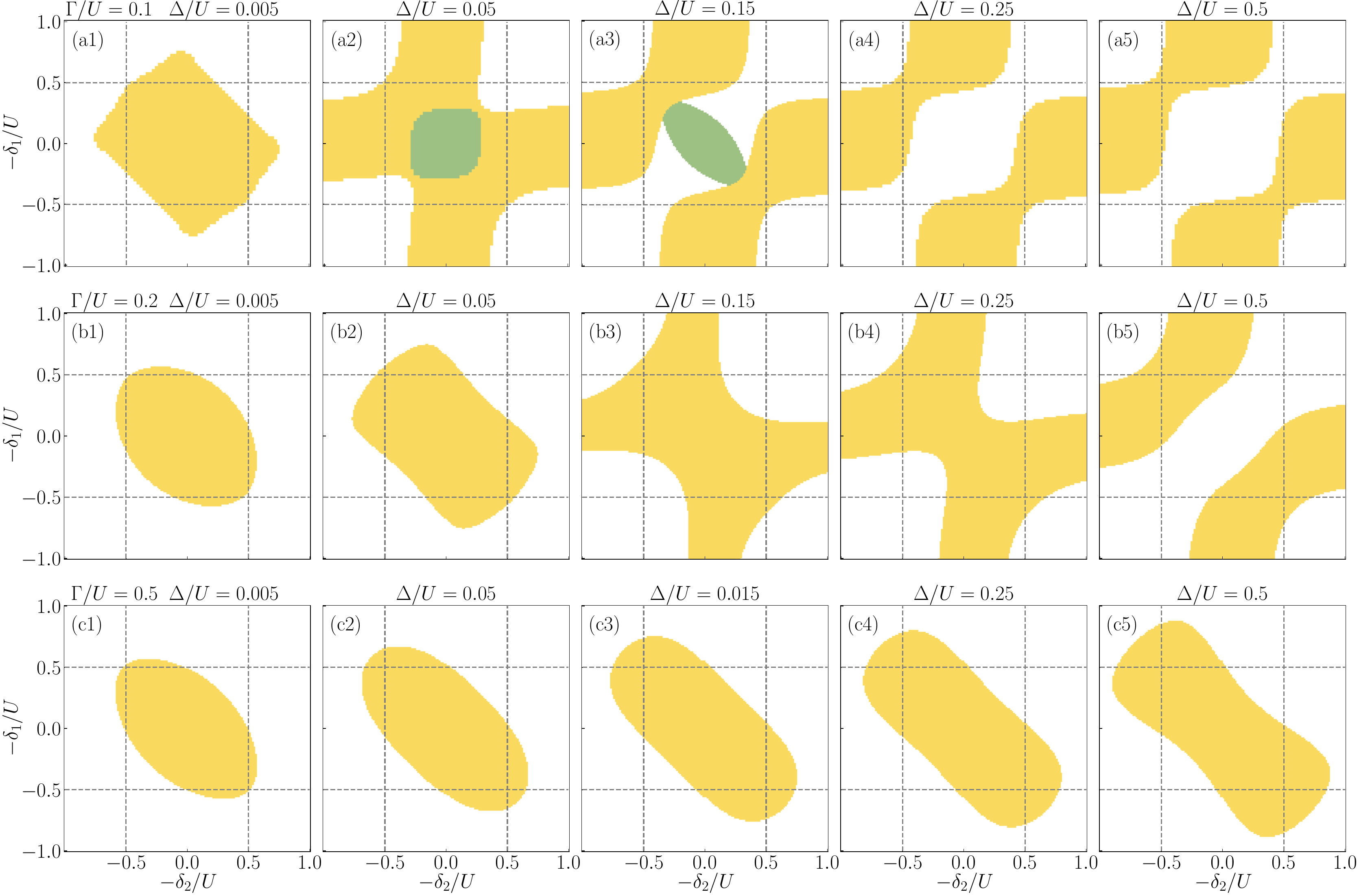}
	\caption{
		The NRG results for the phase diagrams for the coupling symmetric system of two identical parallel QDs at $\nu=1$ as a function of the detuning parameters $-\delta_1/U$ and $-\delta_2/U$
		for varying superconducting gap $\Delta/U$.
		The white, yellow, and green regions correspond to the singlet ($S=0$), doublet ($S=1/2$), and triplet ($S=1$) GS, respectively.
		(a1)-(a5) 
		For $\Gamma/U=0.1$, the evolution with increasing $\Delta/U$ is qualitatively similar to Fig.~\ref{fig:nrg_UG_20},
		where a cross-shaped doublet region splits into separate stripes, and a triplet GS emerges and then disappears as $\Delta/U$ increases.
		(b1)-(b5) 
		For $\Gamma/U=0.2$, the Kondo screening effects dominate over the lead-mediated correlations,
		preventing the formation of a triplet GS for any value of $\Delta/U$.
		(c1)-(c5) 
		At a stronger coupling of $\Gamma/U=0.5$, the Kondo screening processes dominate the behavior,
		resulting in phase diagrams with only centrally-positioned doublet region immersed in a singlet GS area.
		The other parameters are the same as in Fig.~\ref{fig:nrg_UG_20}.
		\label{fig:nrg_UG}}
\end{figure*}

\begin{figure*}
	\includegraphics[width=1.0\textwidth]{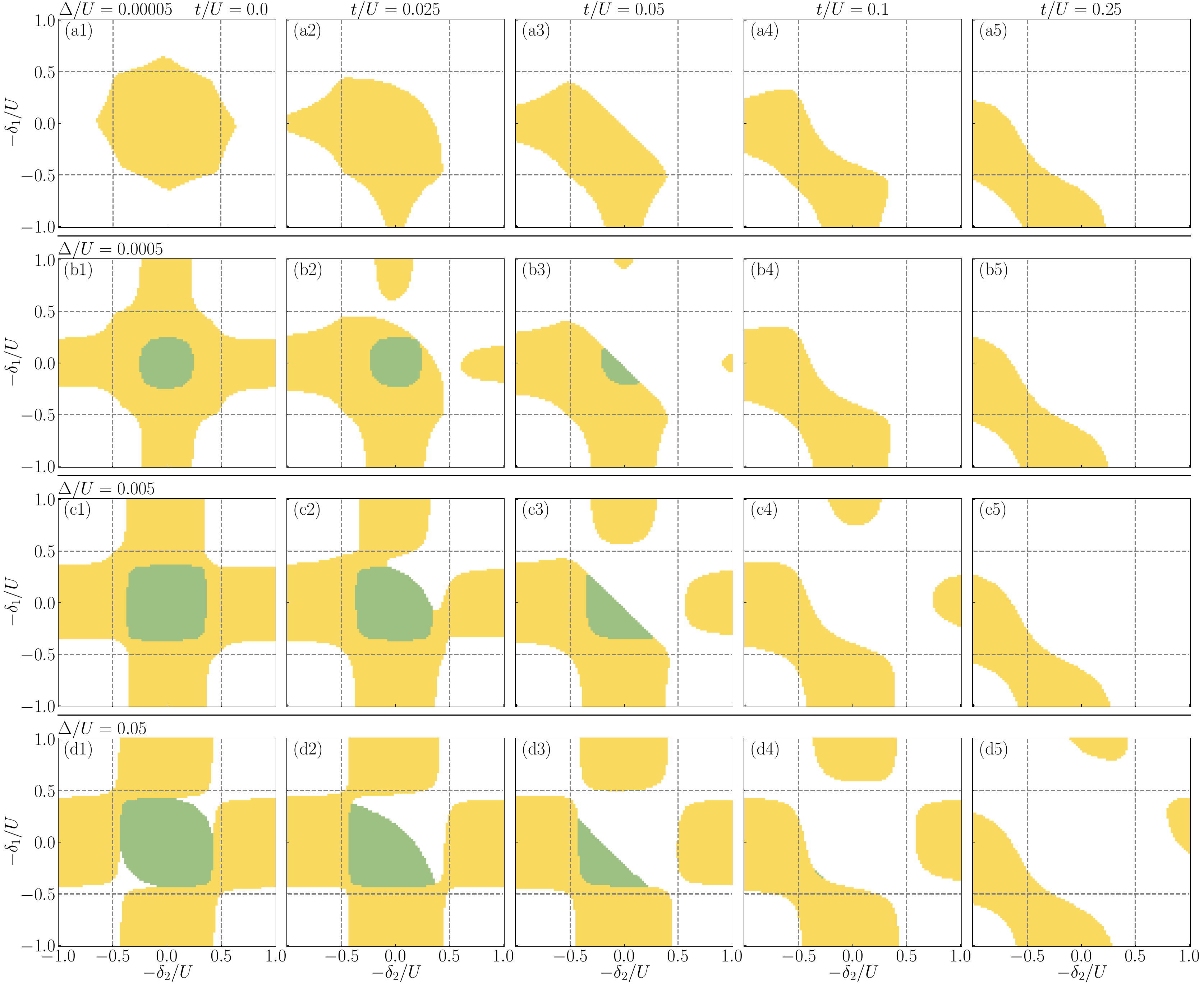}
	\caption{
		The NRG results for the phase diagrams for the coupling symmetric system of two identical parallel QDs at $\nu=1$ calculated as a function of the detuning parameters $-\delta_1/U$ and $-\delta_2/U$
		for varying inter-dot hopping strength $t/U$ for $\Gamma /U = 0.05$.
		The white, yellow, and green regions correspond to singlet (S=0), doublet (S=1/2), and triplet (S=1) GS, respectively.
		The rows correspond to different values of the superconducting energy gap (a) $\Delta/U = 5\times10^{-5}$, (b) $5\times10^{-4}$,
		(c) $5\times10^{-3}$ and (d) $5\times10^{-2}$.
		(a1)-(a5) 
		Initially, with the interdot hopping set to $t/U=0$, the phase diagrams exhibit a doublet GS island,
		which gets distorted as $t/U$ increases.
		(b1)-(b5), (c1)-(c5), (d1)-(d5) 
		For larger values of $\Delta/U$, a triplet GS emerges initially that shrinks and disappears entirely, as $t/U$ is increased.
		Finally, one stripe of the doublet phase remains immersed in the singlet GS region.
		For$t/U \neq 0$, the emerging asymmetry across the diagonal $\delta_1=-\delta_2$,
		is governed by the ph-transformation $\mathcal{T}_1$ discussed in Sec.~\ref{sec:app_ph}.
		Except of $t/U$, all parameters are the same as in Fig.~\ref{fig:nrg_UG_20}.
		\label{fig:nrg_t}
	}
\end{figure*}

When a parallel DQD system is connected to a single SC lead, the competition of superconductivity and screening is accompanied by significant lead-mediated interactions between the constituent QDs, even if no direct tunneling between the dots is present. As follows from Eq.~\eqref{eq:Sigma_resolved}, the off-diagonal blocks of the self-energy, which encode lead-mediated interactions between the QDs, are proportional to $\sqrt{\Gamma_1 \Gamma_2}$ at $\nu=1$. Thus, even in highly asymmetric coupling scenarios, they cannot be rendered marginal with respect to both direct tunnelings simultaneously. 

Let us therefore focus on the scenario where crossed Andreev reflections and their direct counterparts are of the same magnitude. This can be achieved in a coupling-symmetric scenario $\Gamma_1 = \Gamma_2 \equiv \Gamma$ with two identical dots $U_1 = U_2 \equiv U$ at $\nu=1$ as discussed in Sec.~\ref{sec:self_energy}. To showcase the diverse phenomenology possible under such conditions, we first select $\Gamma/U = 0.05$ and vary the gap parameter $\Delta/U$. 
The obtained series of phase diagrams is shown in Fig.~\ref{fig:nrg_UG_20}. The same procedure is then repeated with a larger step in $\Delta/U$ for three additional cases of progressively increasing ratio $\Gamma/U$, as presented in Fig.~\ref{fig:nrg_UG}. 

We first examine the $\Gamma/U = 0.05$ case presented in Fig.~\ref{fig:nrg_UG_20}.
When the SC effects are excluded, as in Fig.~\ref{fig:nrg_UG_20}$(a)$ for $\Delta = 0$,
the phase diagram is governed solely by the Kondo screening and RKKY interactions \cite{Eickhoff-2018},
leading to a well-known behavior of a doublet phase in the center of the diagram
occupying almost the entire region of $-1/2 < \delta_1, \delta_2 < 1/2$,
in line with the qualitative picture of Fig.~\ref{fig:models}(c).
The small bendings of the observed parity transition lines are to be attributed to the presence of lead-mediated RKKY interactions for finite coupling $\Gamma$.
The doublet GS in the middle of the phase diagram is associated with the underscreened Kondo effect \cite{Nozieres1980Mar,Posazhennikova2005Jan,Mehta2005Jul,Roch2009Nov},
in which only one of the spins occupying the DQD can be screened,
while the other one is responsible for $S=1/2$ GS.

As the gap opens up, the ABSs form in the DQD due to the SC-proximity effect while the Kondo screening is gradually suppressed.
This causes the central doublet region to expand along the $\delta_1$ and $\delta_2$ axes,
which is illustrated in Figs.~\ref{fig:nrg_UG_20}(b) and (c) for $\Delta/U=5 \times 10^{-5}$ and $\Delta/U=1\times 10^{-4}$, respectively.
Increasing the gap further, introduces a triplet GS (green) in the middle of the phase diagram,
as visible in Fig.~\ref{fig:nrg_UG_20}(d) for $\Delta/U \approx 5 \times 10^{-4}$.
Noteworthy, the triplet GS, which is strictly forbidden in serial DQD arrangements due to antiferromagnetic
interaction triggered by the hopping \cite{Zitko-2015},
could give rise to the triplet blockade in CPS devices even at equilibrium.
The triplet region then continues to grow until $\Delta/U \approx 5 \times 10^{-3}$, which is shown in Fig.~\ref{fig:nrg_UG_20}(e).
Afterwards, it rapidly shrinks, as seen in Figs.~\ref{fig:nrg_UG_20}(f) and (g) for $\Delta/U = 5 \times 10^{-2}$
and $\Delta/U = 1.5 \times 10^{-1}$.
On the other hand, completely disrupting the central triplet area
requires a large ratio of $\Delta/U = 5 \times 10^{-1}$, as shown in Fig.~\ref{fig:nrg_UG_20}(h).

The doublet GS also undergoes significant evolution as $\Delta/U$ increases.
Initially, it retains a cross-shaped form, but already for $\Delta/U = 5 \times 10^{-2}$,
it splits along the main diagonal $\delta_1 = \delta_2$ into two doublet ribbons, see Fig.~\ref{fig:nrg_UG_20}(f).
When the triplet GS disappears at $U/\Delta = 0.5$, only the bending of the doublet GS area
changes as $\Delta/U$ increases while alternating stripes of singlet and doublet GS
elongate further along the main diagonal $\delta_1 = \delta_2$. Once again, this is a distinctive feature to the serial DQD case,
where the stripes follow the diagonal $\delta_1 = -\delta_2$ \cite{Grove-Rasmussen-2018}.

In the previously selected case of $\Gamma/U = 0.05$,
the Kondo temperature at $\Delta=0$ is approximately
equal to $T_K/U \approx 6\cdot 10^{-5}$,
as calculated from the Haldane formula for one metallic channel screening one QD described by $U$ and $\Gamma$ matching the present model \footnote{The Haldane formula of the following form is applied: $T_K = \sqrt{U \Gamma /2} \, \textit{exp} \left[ \pi \varepsilon (\varepsilon + U )/(2 U \Gamma) \right] $}.
In Fig.~\ref{fig:nrg_UG}, we therefore consider three successively larger ratios of $\Gamma/U$,
which correspond to increasing values of the Kondo temperature.
First, at $\Gamma/U = 0.1$, only quantitative changes appear compared to the previous case.
However, for $\Gamma/U = 0.2$, the screening processes become dominant
and completely prevent the formation of the triplet GS.
Despite this, the alternating singlet and doublet stripes pattern
is recovered at $\Delta/U = 0.5$, as shown in Fig.~\ref{fig:nrg_UG}(b5).
Further increasing the coupling to $\Gamma/U = 0.5$
causes the Kondo screening to dominate the parallel DQD system.
The only remaining feature is an island of doublet GS in the middle of the phase diagram,
observed over several orders of magnitude in $\Delta/U$.
The striped behavior does not manifest even at $\Delta/U = 0.5$.

Let us finally note that so far all observed phase diagrams possessed
two symmetric mirror axes $\delta_1=\delta_2$ and $\delta_1=-\delta_2$, respectively.
While the first is a trivial consequence of selecting coupling
symmetric scenario with two identical dots, the latter one is the consequence of
the ph-transformations at $t=0$, as discussed in Sec.~\ref{sec:app_ph}.
However, when $t \neq 0$, this second symmetry will be broken as shown next.

\subsection{Effects of non-zero inter-dot hopping \label{sec:exact_tdep}}

The results of phase diagrams at zero inter-dot hopping $t$ demonstrate a complex behavior of the parallel DQD model due to the influence of lead-mediated correlations. Adding non-zero inter-dor hopping $t$ to the picture,
makes the outcome even more elaborate. For further exploration, we therefore select the most interesting scenario allowing the triplet GS formation. We thus choose $\Gamma/U=0.05$ in the coupling symmetric scenario with two identical dots, particularly examining the cases demonstrated in Figs.~\ref{fig:nrg_UG_20}(b), (d), (e), and (f) for $t=0$. We then vary the inter-dot hopping $t$ over a wide range to observe its effect on the phase diagrams, as shown in Fig.~\ref{fig:nrg_t}. 

Two main effects arise when $t \neq 0$. First, the mirror symmetry around the $\delta_1=-\delta_2$ axis is lost, contrasting thus with the $t=0$ case. As discussed in Sec.~\ref{sec:app_ph}, this is consistent with the action of the ph-transformation $\mathcal{T}_1$, where simultaneous sign reversal of $\delta_1$, $\delta_2$ and $t$ was shown to form a symmetry operation of the phase diagram. Consequently, the symmetry along the $\delta_1=-\delta_2$ axis must be broken when $t \neq 0$. We note that the effect of the asymmetry in Fig.~\ref{fig:nrg_t} is already well pronounced at all presented ratios of $\Delta/U$ for the first non-zero value of $t$, i. e. $t/U=0.025$, selected herein.

The second most important feature is the gradual disruption of the triplet GS as $t/U$ increases. This evolution is expected due to inter-dot hopping mediating fluctuations between the two constituent dots, favoring singlet over triplet GSs. The triplet area starts to shrink and move apart in the quadrant defined by $-\delta_1, -\delta_2 > 0$ when $t > 0$, which is accompanied by the contraction of the doublet area, favoring the singlet GS once again. Eventually, the doublet regions separate then completely as seen in Figs.~\ref{fig:nrg_t}(b3), (c3), and (d3). Finally, at $t/U=0.25$ the doublet GS are completely removed from the depicted portion of the $-\delta_1, -\delta_2 > 0$ quadrant with only the case of $\Delta/U=0.05$ being a notable exception. Contrary, in the quadrant $-\delta_1, -\delta_2 < 0$, a connected stripe of doublet GS remains intact in the area of $-\delta_1, -\delta_2 < 0$.

In summary, while the behavior observed at $t = 0$ is complex with regard to the size of the gap, at sufficiently large $t/U$ it evolves into one common scenario  dominated by the singlet phase that is accompanied by one connected doublet stripe in the region of $\delta_1, \delta_2 > 0$ for $t>0$. The triplet GSs are then completely removed from the phase diagram. This behavior is universal for arbitrary ratios of $\Delta/U$.

\section{Approximate solutions}

\subsection{Limitations of the AL theory \label{sec:al_limits}}

\begin{figure}
\includegraphics[width=1.0\columnwidth]{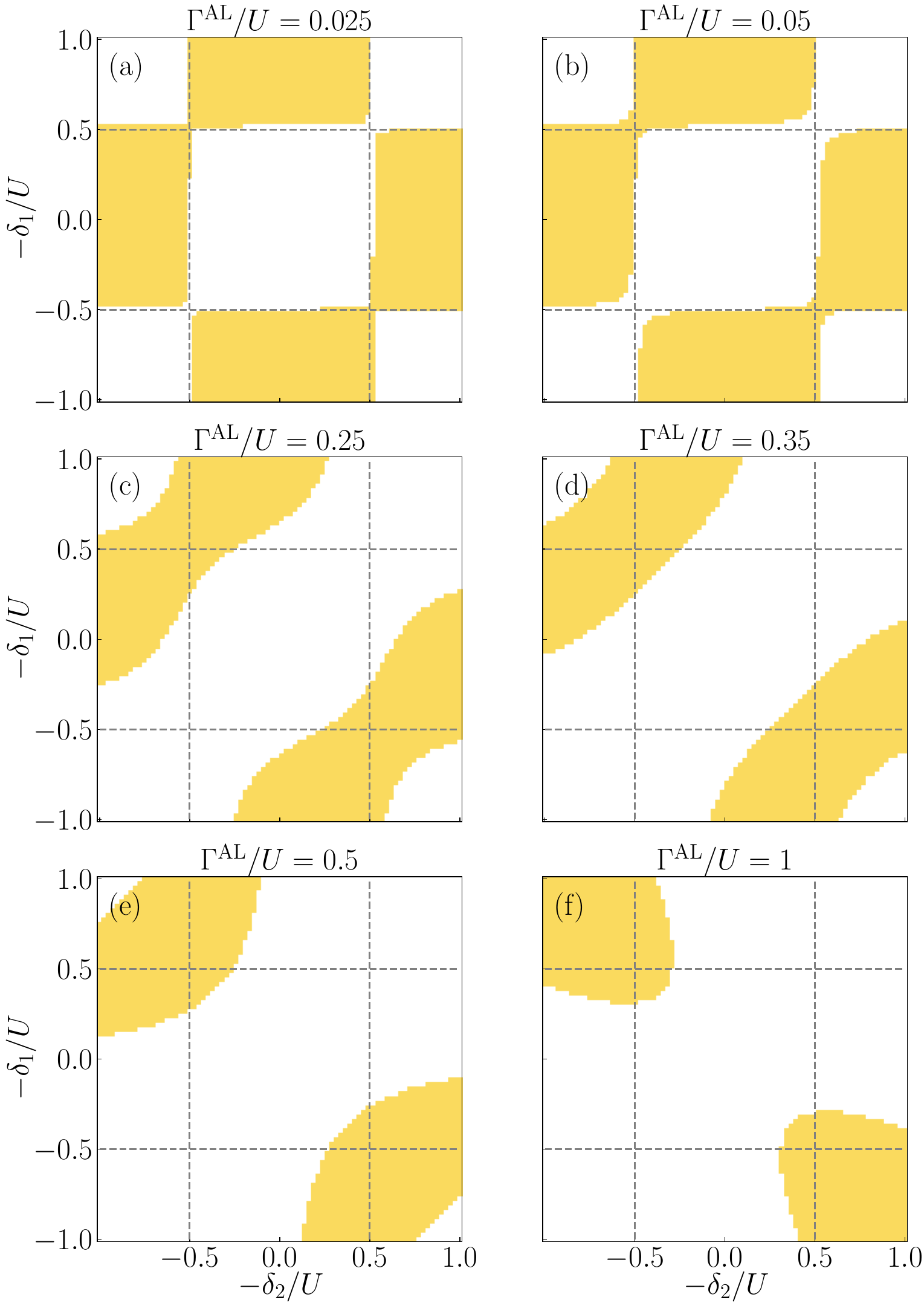}
\caption{ 
	Phase diagrams of two identical parallel dots ($U_1 = U_2 \equiv U$) calculated
	using the AL theory for the symmetric coupling scenario
	with $\Gamma^{\mathrm{AL}}_1 = \Gamma^{\mathrm{AL}}_2 \equiv \Gamma^{\mathrm{AL}}$ at $\nu=1$
	as a function of the detuning parameters, $-\delta_1/U$ and $-\delta_2/U$. At $t=0$, $\Gamma^{\mathrm{AL}}/U$ remains the only relevant scale to be varied in the AL theory.
	Panels (a)-(c) with $\Gamma^{\mathrm{AL}}/U \leq 0.25$ are in a good qualitative agreement
	with the NRG results observed for large $\Delta$
	in Figs.~\ref{fig:nrg_UG_20} and \ref{fig:nrg_UG}.
	However, the remaining panels (d)-(f), with larger $\Gamma^{\mathrm{AL}}/U$ values,
	have no counterpart in the exact results.
	Overall, the AL fails in predicting the triplet GS completely.
	The same also applies to GAL, as it only rescales the parameters that enter the AL theory.
	\label{fig:al}}
\end{figure}

\begin{figure*}
\includegraphics[width=1.0\textwidth]{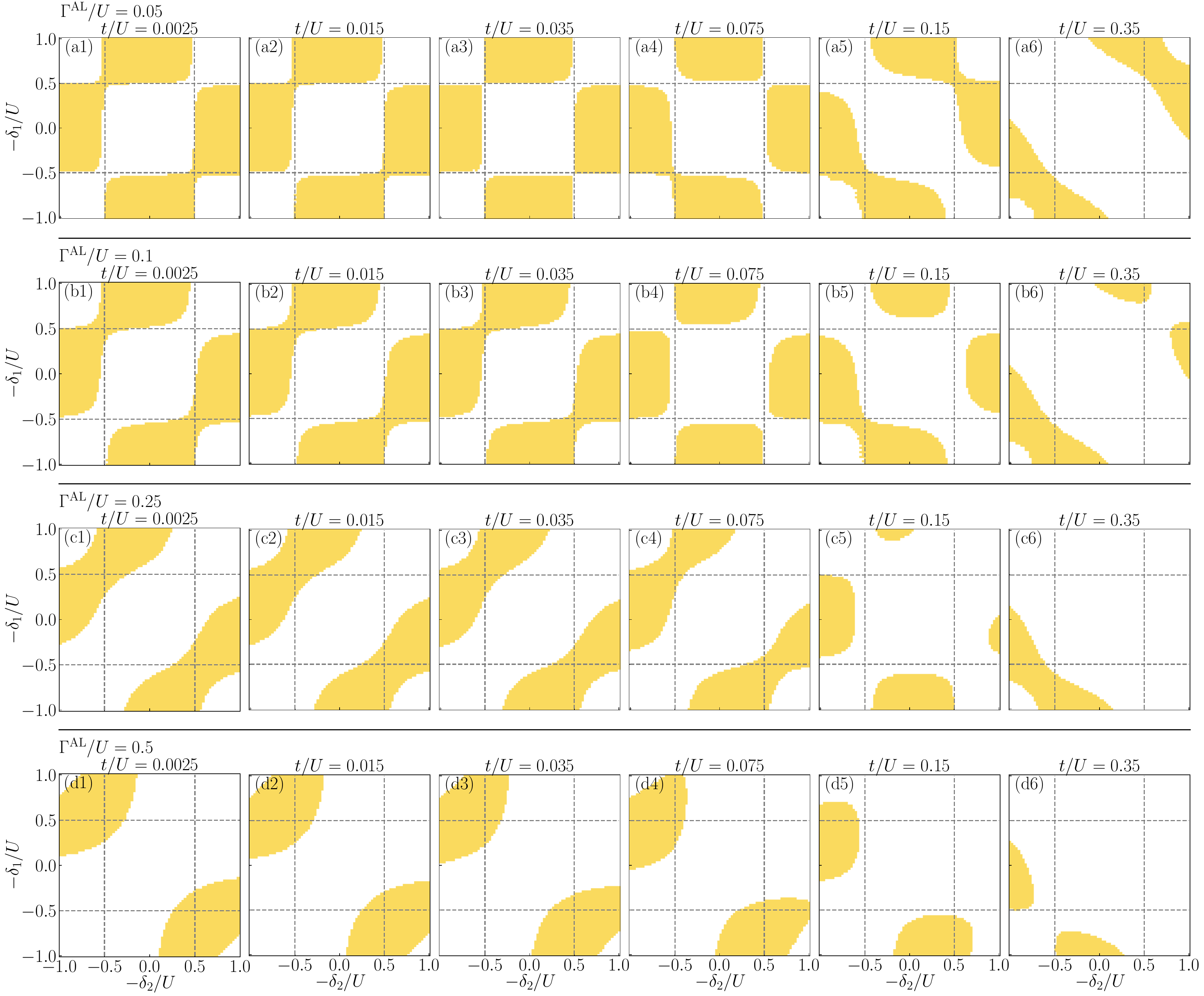}
\caption{ 
	Phase diagrams of two identical parallel dots ($U_1 = U_2 \equiv U$) calculated
	using the AL theory for non-zero inter-dot hopping $t$
	in a symmetric coupling scenario with $\Gamma^{\mathrm{AL}}_1 = \Gamma^{\mathrm{AL}}_2 \equiv \Gamma^{\mathrm{AL}}$
	and $\nu=1$.
	For finite inter-dot hopping, two independent scales, $\Gamma^{\mathrm{AL}}/U$ and $t/U$,
	are present in the AL model.
	A series of panels in one row corresponds to a fixed ratio of $\Gamma^{\mathrm{AL}}/U$, while $t/U$ is varied.
	 Due to $t \neq 0$, the symmetry across the diagonal $\delta_1=-\delta_2$
	 is broken as discussed in Sec.~\ref{sec:app_ph}.
	 However, once again AL, and consequently also GAL, fail to reproduce the triplet GS,
	 as observed by exact methods in Fig.~\ref{fig:nrg_UG},~\ref{fig:nrg_UG_20} and \ref{fig:nrg_t}.
	 At intermediate values of $t/U$, AL predicts erroneous phase diagrams,
	 which have no counterpart in the exact results.
	 Moreover, the behavior for $\Gamma^{\mathrm{AL}}/U=0.5$, as shown in panels (d1)-(d5),
	 is not observed in exact results at all.
	 Consequently, the AL theory, or its generalized form GAL,
	 cannot be extrapolated outside the infinite $\Delta$ limit.
	\label{fig:al_t}
}
\end{figure*}

The AL theory and its generalization GAL are derived in a well-controlled procedure outlined in Sec.~\ref{sec:al},
which suggests their validity being limited to cases,
where $\Delta$ is the largest energy scale in the problem.
Nevertheless, in single QD and serial DQD problems both were shown
to fare reasonably well beyond such a natural expectation.
While AL validity was mainly of qualitative nature,
GAL fixed even the corresponding quantitative predictions \cite{Zonda-2023}.
Therefore, one often sees AL being applied even outside
of such strict limitations when more complex systems are addressed.
In this section, we will however reveal that AL and consequently also GAL
fail in the present case even in qualitative terms.

\begin{figure*}
	\includegraphics[width=1.0\textwidth]{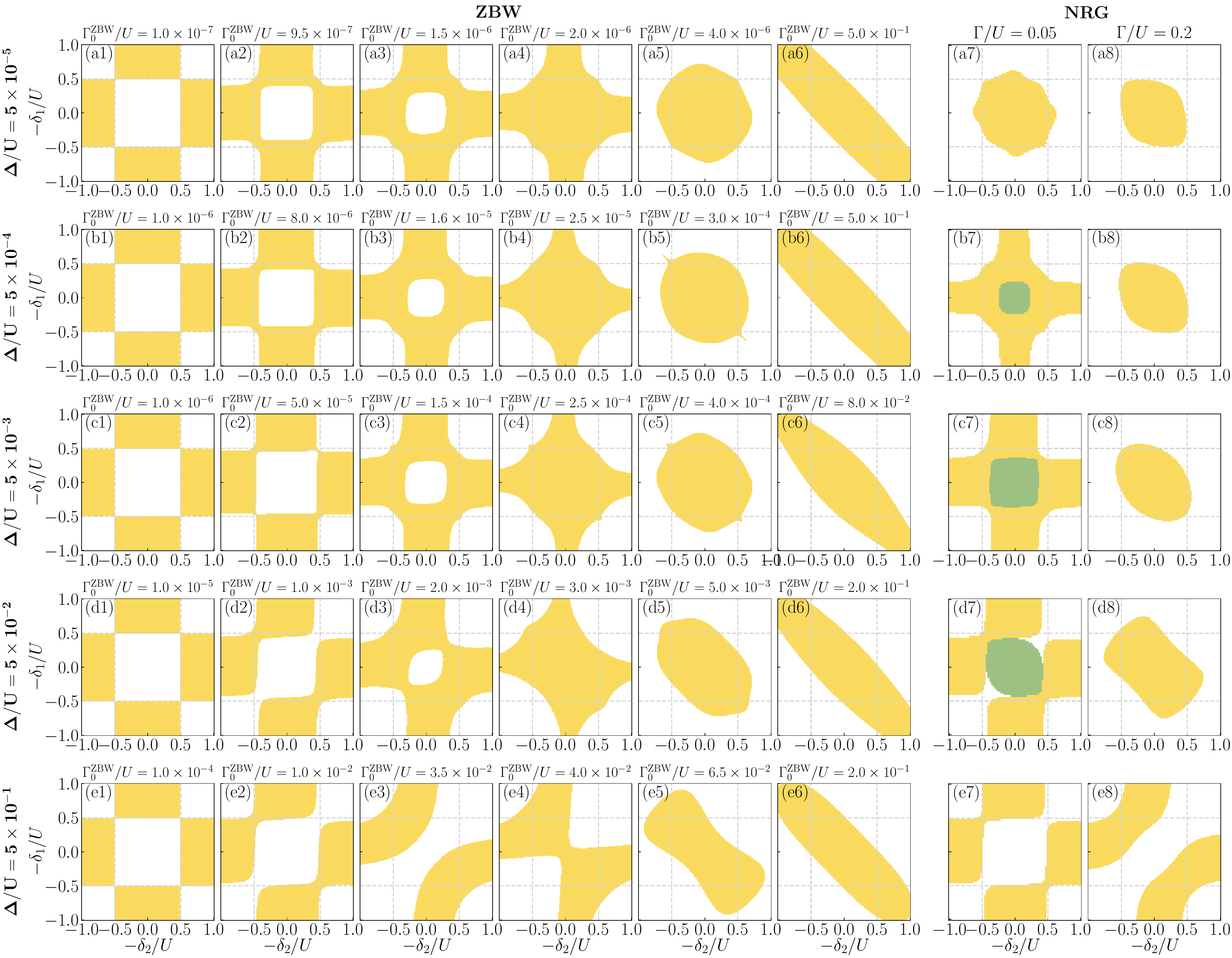}
	\caption{ 
		Comparison of the phase diagrams obtained using the ZBW approximation
		and the numerically exact NRG method for the coupling symmetric ($\Gamma^{\mathrm{ZBW}}_{1,0}=\Gamma^{\mathrm{ZBW}}_{2,0}\equiv\Gamma^{\mathrm{ZBW}}_0$) system of two identical QDs ($U_1=U_2 \equiv U$) at $\nu=1$.
		The first six columns, show the phase diagrams calculated within the ZBW approximation for different ratios of $\Gamma^{\mathrm{ZBW}}_0/U$ and $\Delta/U$,
		where $\Gamma^{\mathrm{ZBW}}_0$ is the effective hybridization within the ZBW approximation.
		The last two columns present
		the corresponding NRG results for two selected ratios of $\Gamma/U$,
		which serve as the exact benchmark.
		Each row corresponds to a fixed ratio of $\Delta/U$,
		ranging from $5 \times 10^{-5}$ (top row) to $5 \times 10^{-1}$ (bottom row), as indicated.
		The ZBW approximation fails to capture the triplet phase,
		which excludes it from modeling a significant portion of the parameter space for the present system.
		The parameters for NRG calculations are the same as in Fig.~\ref{fig:nrg_UG_20}.
		\label{fig:zbw}}
\end{figure*}

To this end, we again start with the $t=0$ case in the coupling-symmetric scenario $\Gamma^{\mathrm{AL}}_1=\Gamma^{\mathrm{AL}}_2 \equiv \Gamma^{\mathrm{AL}}$ at $\nu=1$ with identical dots $U_1=U_2 \equiv U$. Here, the crossed and direct Andreev reflections are of the same order. Since $\Delta$ is not present in AL, only $U$ and $\Gamma^{\mathrm{AL}}$
remain as independent parameters for $t=0$.
Selecting $U$ as the energy unit, leaves us therefore with the ratio $\Gamma^{\mathrm{AL}}/U$
as being a free parameter of the AL theory at $t=0$.
The resulting phase diagrams are then shown in Fig.~\ref{fig:al}.

In detail, when $\Gamma^{\mathrm{AL}}/U=0.025$, as in Fig.~\ref{fig:al}(a), an almost perfect checkerboard pattern of singlet and doublet phases with avoided crossings along the main diagonal $\delta_1=\delta_2$ appears. This is a clear sign of lead-mediated interactions in the parallel DQD scenario,
as discussed already in the case of Figs.~\ref{fig:models}(e)-(f). When $\Gamma^{\mathrm{AL}}/U$ is increased, the region of avoided crossings grows, as demonstrated in Fig.~\ref{fig:al}(b) for $\Gamma^{\mathrm{AL}}/U=0.05$. Upon further increasing $\Gamma^{\mathrm{AL}}/U$,
the distance between the doublet regions grows, as can be seen in Figs.~\ref{fig:al}(c) and (d)
for $\Gamma^{\mathrm{AL}}/U=0.25$ and $\Gamma^{\mathrm{AL}}/U=0.35$, respectively.
Increasing the ratio up to $\Gamma^{\mathrm{AL}}/U=1$, causes
the stripe pattern to collapse, as shown in Figs.~\ref{fig:al}(e) and (f),
and the doublet regions extend along the diagonal $\delta_1=-\delta_2$.
However, this behavior is not observed
in exactly calculated phase diagrams for realistic parameters with finite $\Delta$.

Obviously, the AL theory completely misses the triplet GSs
and only the striped scenarios can be reliably modeled,
which is only relevant for $\Delta \gtrsim U$
by simultaneously ensuring $\Delta > \Gamma$.
In other words, for the present case, it is impossible to extrapolate
AL beyond its natural limit of $\Delta$ being the dominant energy scale in the system.
Consequently, for $t=0$, AL and also the corresponding GAL theory
fail completely in predicting triplet phases
and fall also short of describing the complex evolution of the doublet phase,
except for the checkerboard and stripe patterns.

For finite hopping between the quantum dots, $t \neq 0$,
the situation does not improve either, as demonstrated in Fig.~\ref{fig:al_t}.
Now, the AL model possesses two scales, namely $\Gamma^{\mathrm{AL}}/U$ and $t/U$ that need to be scanned.
We have therefore assumed four values of the ratio $\Gamma^{\mathrm{AL}}/U \in \{ 0.05, 0.1, 0.25, 0.5 \}$,
and then gradually increased the remaining ratio of $t/U$. The obtained phase diagrams respect the action of transformations $\mathcal{T}_1$ and $\mathcal{T}_2$
as discussed in Sec.~\ref{sec:app_ph} similarly to the numerically exact NRG data. In detail, the 
symmetry along the diagonal $\delta_1=-\delta_2$ is broken for $t\neq 0$.
However, unlike in exact results the triplet GS is not observed at all. Thus, only qualitative behavior of the doublet stripes
being disintegrated can be captured by AL when $t \neq 0$.

Altogether, the atomic limit theory, and its counterpart GAL,
are subject to severe restrictions if a correct behavior
of a realistic parallel DQD system is required to be modeled.
These restrictions align with the $\Delta \rightarrow \infty$ nature of the approximation
and should be observed strictly in the present DQD model.

\begin{figure*}
	\includegraphics[width=1.0\textwidth]{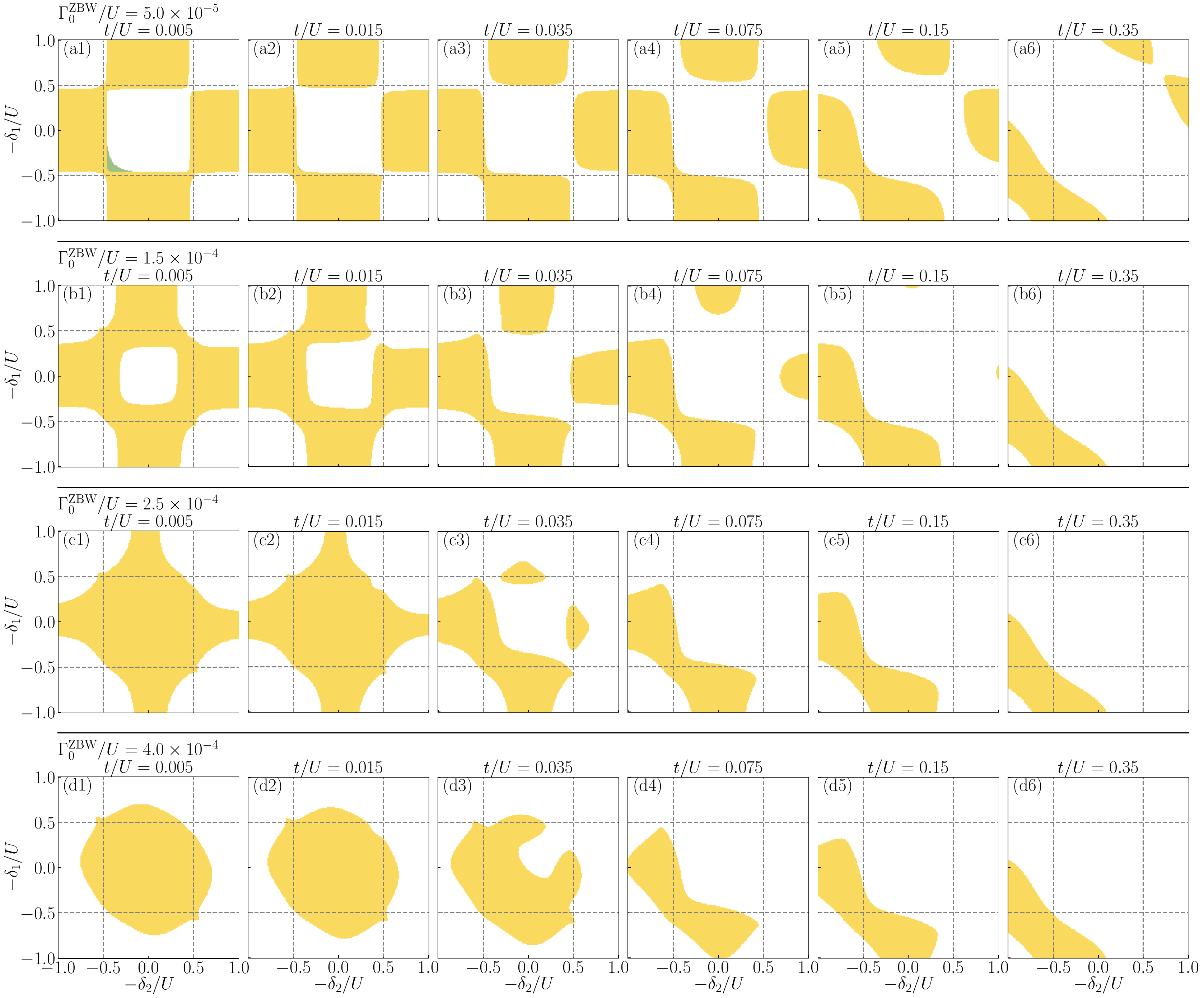}
	\caption{ 
		Phase diagrams for two identical parallel dots ($U_1 = U_2 \equiv U$) calculated
		using the ZBW approximation for the symmetric coupling scenario
		with $\Gamma^{\mathrm{ZBW}}_{1,0}=\Gamma^{\mathrm{ZBW}}_{2,0}=\Gamma^{\mathrm{ZBW}}_0$ at $\nu=1$.
		Four cases from Figs.~\ref{fig:zbw}(c2)-(c5) with $\Delta/U=5 \times 10^{-3}$ have been selected. 
	    Each row shows results for fixed $\Gamma^{\mathrm{ZBW}}_0/U \in \{ 5.0 \times 10^{-5}, 1.5 \times 10^{-4}, 2.5 \times 10^{-4}, 4.0 \times 10^{-4}\}$, whereas the ratio of $t/U$ increases horizontally, as indicated. 
	    While at initial stages the phase diagrams undergo complex development, they all evolve into a regime dominated by singlet GS with one or two separated doublet regions. 
		\label{fig:zbw_t}}
\end{figure*}

\subsection{Failure of the ZBW approximation \label{sec:ZBW_fail} }

When the zero bandwidth approximation is employed to effectively model the system,
values of hybridization parameters need to be adjusted to match the exact or experimental phase diagrams as successfully
undertaken for single QD and serial DQDs~\cite{Grove-Rasmussen-2018}. Moreover,
due to its simplicity the ZBW approximation has been widely extrapolated to other more complex models where exact benchmarks were not performed \cite{vonOppen-2021}.
However, by thoroughly analyzing the present DQD setup
we clearly demonstrate that ZBW does not represent a universal tool
for modeling QD-based superconducting systems,
as it fails even in this relatively simple scenario. 

To this end, we once again consider identical dots $U_1=U_2 \equiv U$ in the coupling-symmetric scenario $\Gamma^{\mathrm{ZBW}}_{1,0}=\Gamma^{\mathrm{ZBW}}_{2,0} \equiv \Gamma^{\mathrm{ZBW}}_0$ 
with $\nu=1$.
Using the Coulomb repulsion $U$ as the energy unit,
leaves then three independent ratios of $\Delta/U$, $t/U$, and $\Gamma^{\mathrm{ZBW}}_0/U$
to be freely adjusted in the corresponding ZBW Hamiltonian~\eqref{eq:zbwH}.
Focusing initially on the simpler $t=0$ scenario,
where the inter-dot interactions are mediated solely by the common BCS lead,
we fix the ratio of $\Delta/U \in \{ 10^{-1}, 10^{-2}, 10^{-3}, 10^{-4}, 10^{-5} \}$,
which span five orders of magnitude. 

The strategy is then to vary $\Gamma^{\mathrm{ZBW}}_0/U$,
which is the only remaining free parameter in ZBW for $t=0$,
in a hope to find a possible value, 
which replicates the corresponding exact results from Sec.~\ref{sec:exact_t0}.
The systematic scan is shown in Fig.~\ref{fig:zbw}.
Each row presents the results for the same ratio of $\Delta/U$ with first six columns
representing the ZBW outcome, while two exactly obtained (by NRG) phase diagrams
are shown afterwards for $\Gamma/U=0.05$ and $\Gamma/U=0.2$, respectively.
The ZBW results have been carefully selected to represent
all qualitative regimes of the ZBW approximation in the case of  $t=0$.

Starting with the $\Gamma/U=0.2$ case, one could argue that the ZBW approximation
might be used to model at least some gross features of the parallel DQD system,
as the doublet GS island is always found within the ZBW solutions.
For example, when $\Delta/U=10^{-3}$, the best qualitative fit could be attributed to Fig.~\ref{fig:zbw}(c5).
While NRG predicts the doublet region always within the bounds
of $|\delta_1|, |\delta_2| \leq 1/2$, ZBW predicts always a much larger extent,
with only a minor adjustment obtained by better fitting $\Gamma^{\mathrm{ZBW}}_0$.
This is associated with the fact that NRG results include all
coupling-induced renormalization effects that are basically absent in the ZBW approximation.
Similarly, also the striped phase observed in Fig.~\ref{fig:zbw}(e8) might
be qualitatively well matched with a ZBW solution of Fig.~\ref{fig:zbw}(e3).
In this case, the match is obviously much better,
making the striped regime suitable for simplistic modeling.

However, going further to the case of $\Gamma/U=0.05$, reveals
ZBW to fail completely even in qualitative terms, as triplet GSs are not predicted by ZBW approximation at all
for a broad range of $\Delta/U$ spanning approximately the interval of $\left[10^{-3},10^{-1}\right]$. On the other hand, the shape of the singlet and doublet regions
away from the central portion of the phase diagram
seem to match the exact results if appropriate values of $\Gamma^{\mathrm{ZBW}}_0$ are selected.
For instance, at $\Delta/U=5 \times 10^{-3}$, there seem to be a relatively
good match between Figs.~\ref{fig:zbw}(c3) and (c7) in terms of singlet and doublet phases. 

ZBW thus fails at $t=0$ completely in predicting the triplet phases,
while for doublet and singlet phases it returns results,
which capture only very general features, 
except of the parallel DQD system being in the striped regime,
as observed when $\Delta/U$ is the dominant scale in the system.
Consequently, for $t=0$, the ZBW approximation experiences the same severe
limitations as the AL theory discussed previously.
The failure of ZBW to capture the triplet GS
is thus inherently linked to lead-mediated interactions between the dots.
Such interactions were absent in the serial DQD system,
potentially leading to an erroneous belief that ZBW
is a general tool for modeling DQD systems.
However, it also suggests that with direct inter-dot interactions introduced in the parallel scenario,
the ZBW approximation might improve.

To examine this matter, we have therefore
selected four exemplary cases from Figs.~\ref{fig:zbw}(c2)-(c5)
and varied $t$ in a sufficiently large range of $t/U$,
with the results shown in Fig.~\ref{fig:zbw_t}.
Focusing first on the largest ratio of $t/U$ in Figs.~\ref{fig:zbw_t}(a6), (b6), (c6) and (d6),
we indeed see that even within the ZBW approximation
the system tends to form a single doublet stripe,
while the triplet phase completely vanishes,
which resembles the behavior observed already in the exact results of Fig.~\ref{fig:nrg_t} and in the much simpler AL approach presented in Fig.~\ref{fig:al_t}. However, ZBW  underestimates the extent of the doublet stripe when compared to the exact results.


In the intermediate regimes observed in the first five columns of Fig.~\ref{fig:zbw_t}, the quality of ZBW's predictions suffers even more substantial inaccuracies,
as the lead-mediated interactions are not negligible in the exact scenario. Consequently, ZBW fails to predict the competition with direct inter-dot hopping
and results therefore in phase diagrams almost completely free of the triplet GS with only Fig.~\ref{fig:zbw_t}(a1) showing a very small triplet area.

Consequently, introducing non-zero $t$ somewhat improves the applicability of ZBW,
but still it is either completely limited to striped regimes,
covered already by AL theory very well, or requires $t$ to be the dominant scale, which actually drives the parallel DQD system into a regime where lead-mediated interactions are not essential.

\subsection{Extended ZBW model \label{sec:eZBW}}

\begin{figure}
	\includegraphics[width=1\columnwidth]{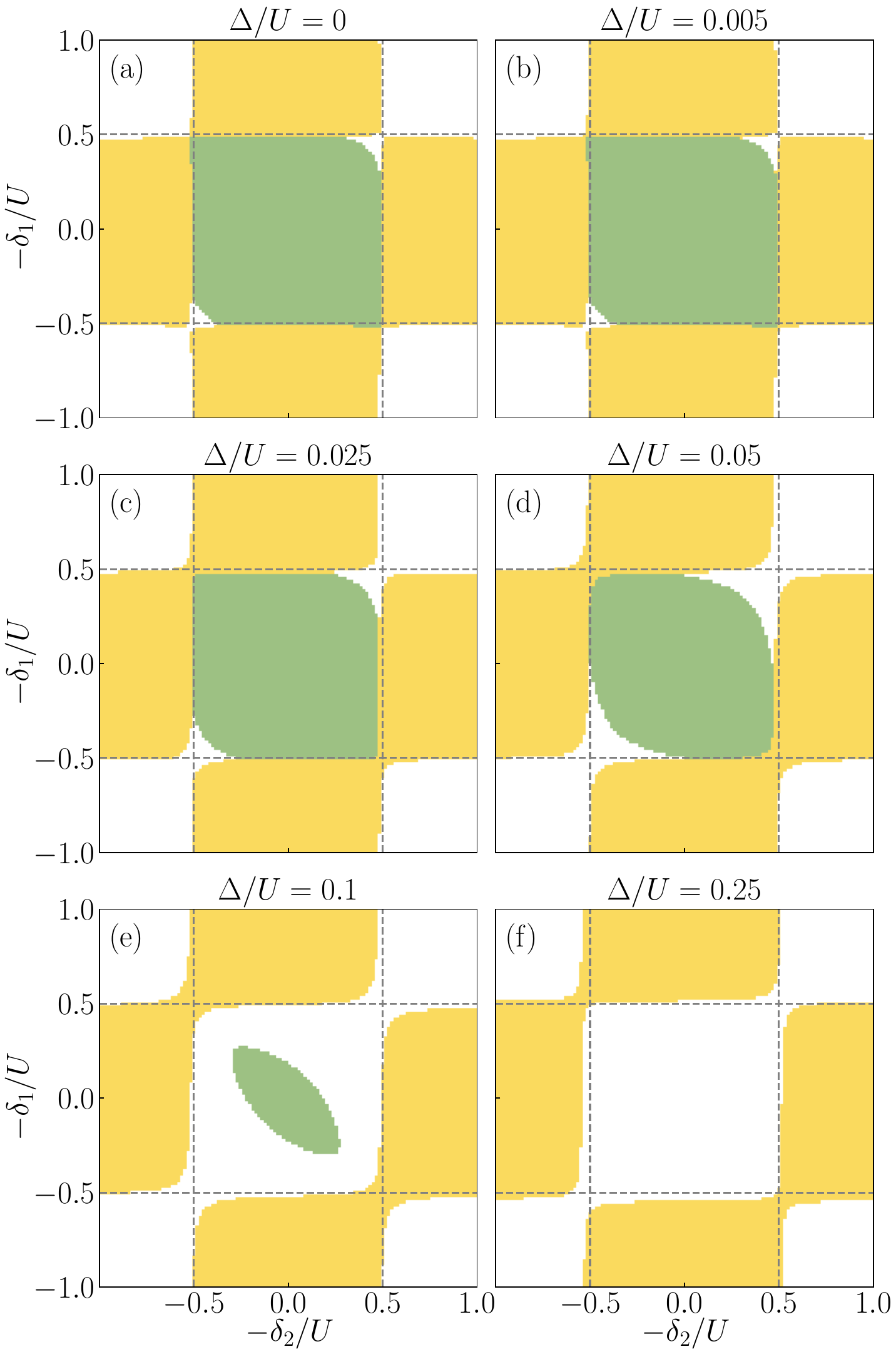}
	\caption{ 
		The phase diagrams for the coupling symmetric ($V^{\mathrm{ZBW}}_{1,0}=V^{\mathrm{ZBW}}_{2,0} \equiv V^{\mathrm{ZBW}}_0$) DQD system calculated via eZBW. Without inter-dot hopping only two independent scales, $V^{\mathrm{ZBW}}_0/U$ and $V^{\mathrm{ZBW}}_1/U$,
		are present in eZBW at given $\Delta/U$. Selecting heuristically the condition $V^{\mathrm{ZBW}}_0/U = V^{\mathrm{ZBW}}_1/U$ we then set both scales to $0.1$ and vary $\Delta/U$.
		(a) 
		In the absence of superconductivity ($\Delta=0$), the triplet GS in the middle of the phase diagram is falsely predicted. This is due to both scales $V^{\mathrm{ZBW}}_0/U$, $V^{\mathrm{ZBW}}_1/U$ being set too large. 
		(b)-(c) 
		As $\Delta/U$ increases, the region of triplet GS shrinks, but it is still
		too large compared to the exact results at given $\Delta/U$.
		(d)-(f)
		The evolution of the phase diagrams is in qualitative agreement with the excat NRG results. At $\Delta/U = 0.25$, the triplet GS disappears and a striped pattern of singlet and doublet phases
		appears in the direction of the main diagonal $\delta_1=\delta_2$.
		\label{fig:ezbw}}
\end{figure}

In previous sections we have demonstrated that both AL-based models and the ZBW approximation can lead
to unphysical predictions for the parallel DQD systems. We therefore emphasize that a good practice is to always
benchmark simplified descriptions of QD-based problems with exact approaches like NRG when extending them
beyond their proven applicability domain. On the other hand, simplified models
offer valuable tools when their limitations are respected. For example, even when constrained to a specific
regime, these models can provide analytical insights that aid in the design of new devices, as illustrated by
the case of equilibrium quartet superconductivity \cite{Chirolli-2024}. 

As an example of such a simplified approach, we now aim to extend the ZBW model to capture the singlet-triplet transitions identified in Sec,~\ref{sec:app_nrg} by exact NRG calculations. Our focus lies in this particular aspect, and we stress that no rigorous claims are made regarding the broader applicability beyond the current model or the effect we aim to describe. Given the importance of lead-mediated interactions in the DQD system, it is natural to extend the ZBW approximation defined in Sec.~\ref{sec:zbw} by adding another SC orbital labeled $1$. It is supposed to couple via a new effective parameter, the hybridization term $V^{\mathrm{ZBW}}_1$, only to the original ZBW orbital labeled $0$. We refer to this approximation as the extended ZBW (eZBW) model.

In eZBW, the parametric space has been further enlarged.
In order to manage such a challenge, we first assume $V^{\mathrm{ZBW}}_1$ to be the same for all level
detunings $\delta_1$ and $\delta_2$, while $\Delta$, $U_1$ and $U_2$ remain equal to their bare values as before.
This leaves $V^{\mathrm{ZBW}}_0$, $V^{\mathrm{ZBW}}_1$, and possibly $t$, to be adjusted
within eZBW. Despite these many simplifying assumptions,
analysis based on eZBW remains still multi-parametric\footnote{The initial values might be inputed from the corresponding surrogate calculations \cite{Baran-2024,Baran-2024-haldane}. However, further adjustments are still required.}. 

To describe the challenging scenario of singlet-triplet transitions, we select the coupling-symmetric scenario at $\nu=1$ and identical dots ($U_1=U_2=U$) without inter-dot interaction. 
We then address the DQD parameters of Fig.~\ref{fig:nrg_UG_20} with $\Gamma/U=0.05$ and select two cases: $\Delta/U = 5 \times 10^{-4}$ and $\Delta/U = 5 \times 10^{-2}$, which have
previously been inaccessible by ZBW or AL-based models. We then vary $V^{\mathrm{ZBW}}_0/U$ and $V^{\mathrm{ZBW}}_1/U$ across several orders of magnitude to analyze the corresponding phase diagrams.
Since the resulting data is rather elaborate, 
we show the complete results in the Appendix in Figs.~\ref{fig:ezbw_0.05} and \ref{fig:ezbw_0.0005}
with panels ordered by increasing $V^{\mathrm{ZBW}}_0$ horizontally and $V^{\mathrm{ZBW}}_1$ vertically.

Starting with the $\Delta/U = 5 \times 10^{-2}$ case,
we notice that a significant portion of the eZBW-parametric space 
matches the exact phase diagram observed in Fig.~\ref{fig:nrg_UG_20}(f) quite well. Selecting  Fig.~\ref{fig:ezbw_0.05}(c3) as the best fit yields then $V^{\mathrm{ZBW}}_0/U=0.0125$ and $V^{\mathrm{ZBW}}_1/U= 0.02$.
Both hybridization terms are thus of similar magnitude,
with $V^{\mathrm{ZBW}}_0$ giving a tunneling rate of the order of $\Gamma^{\mathrm{ZBW}}_0/U \approx 10^{-4}$.
Reducing then the BCS gap further to $\Delta/U = 5 \times 10^{-4}$ once again positively demonstrates eZBW's capabilities in reproducing the exact phase diagram shown in
Fig.~\ref{fig:nrg_UG_20}(d). However, the effective tunneling rates inserted into the eZBW drop even further to $\Gamma^{\mathrm{ZBW}}_0/U \approx 10^{-6}$ as Fig.~\ref{fig:ezbw_0.0005}(d3) with $V^{\mathrm{ZBW}}_0/U=0.001$ and $V^{\mathrm{ZBW}}_1/U=5 \times 10^{-5}$ represents the best fit to the exact result of Fig.~\ref{fig:nrg_UG}(g).

The analysis of the two cases shows that determining both $V^{\mathrm{ZBW}}_0$ and $V^{\mathrm{ZBW}}_1$ is relatively costly even at $t=0$. To avoid such problems we resort to the next simplification and assume both parameters to be equal. We fix their values to
$V^{\mathrm{ZBW}}_0/U=V^{\mathrm{ZBW}}_1/U=0.1$ leaving in analogy to Fig.~\ref{fig:nrg_UG_20} only the ratio $\Delta/U$ to be tuned.
The resulting phase diagrams are shown in Fig.~\ref{fig:ezbw}. Due to the assumption $V^{\mathrm{ZBW}}_0=V^{\mathrm{ZBW}}_1$, unphysical behavior appears in Fig.~\ref{fig:ezbw}(a) at $\Delta=0$ and in Fig.~\ref{fig:ezbw}(b) at $\Delta=0.005$ where Kondo effects dominate the system. Here, the simplified eZBW 
incorrectly predicts a sizable triplet GS in the middle of the phase diagram. Only as $\Delta/U$ is further increased, the triplet GS evolution becomes qualitatively comparable to the exact NRG results obtained earlier, albeit larger values of $\Delta/U$ are required. 



Overall, eZBW allowed us to model singlet-triplet transitions of the full problem. 
However, the rapid evolution of effective tunneling rates as $\Delta/U$ decreases
necessitates extensive parametric scans, diminishing eZBW's practical benefits.
Approximate prescription fixing $V^{\mathrm{ZBW}}_0=V^{\mathrm{ZBW}}_1$ rectified this problem, but is qualitatively applicable only in a specific range of values illustrated in Fig.~\ref{fig:ezbw}(c)-(f). While eZBW shows potential for simplified modeling of the present system, we emphasize that a further quantitative analysis would be required.

\section{Conclusions \label{sec:conslusions} }

In this work, we have systematically investigated the phase diagram of a parallel DQD Andreev molecule using NRG calculations in a symmetric set-up which maximizes lead-mediated interactions. Our key findings reveal a rich phenomenology, including singlet, doublet, and triplet GSs, arising from the intricate interplay between Kondo screening, superconductivity, and lead-mediated interactions between the QDs. Notably, the presence of a triplet GS is a distinct feature of the parallel arrangement with crossed and direct Andreev reflections of the comparable magnitude. This enables strong lead-mediated interactions, which are completely absent in serial DQD setups. The observed triplet GS is of practical importance, for example, in CPS devices, where it might be used to engineer triplet blockades in equilibrium, an effect that was previously predicted only at finite voltages applied to the system.


In the coupling-symmetric scenario with two identical dots ($U_1=U_2\equiv U$) and a set-up maximizing lead-mediated interactions ($\nu=1$), the evolution of the phase diagram depends significantly on the ratios of $\Delta/U$, $\Gamma/U$, and $t/U$, as demonstrated by our exact NRG calculations. The lead-mediated interactions, particularly pronounced in such configurations, proved to be an insurmountable obstacle for the application of effective models, such as the (G)AL or ZBW approximation, as they fail to capture the intricate features of the parallel DQD configuration, even qualitatively. These effective models are therefore constrained only to the scenario where the BCS gap parameter is the dominating scale in the system. While this limitation might be expected for AL and its generalization GAL due to their $\Delta \rightarrow \infty$ character, for ZBW, no such constraints have been observed so far. To clearly understand such limitations is of utter importance, since ZBW has gained popularity in studying of multi-orbital systems on superconducting surfaces. In Ref.~\cite{vonOppen-2021}, the ZBW approximation represented then the basis for the theoretical understanding of spin $S>1/2$ systems. While it has been successfully applied also in real experimental scenarios \cite{Liebhaber-2022,Wang-2024-bulletin}, the present results suggest that unphysical predictions of ZBW are nevertheless possible in multi-orbital systems. 

On the other hand, extending ZBW by adding another SC orbital, in the similar fashion as in Ref.~\cite{Pavesic-2024}, has proven to largely mitigate the deficiencies of the other effective models in our initial analysis based solely on the GS properties of the parallel DQD system. Further research regarding ABSs, thermodynamic quantities and eventual (analytic) prescription of the effective values is, nevertheless, still required.

Our findings underscore the need for caution when extrapolating the use of effective models beyond single QD and serial DQD systems, where AL-based models and ZBW have been exclusively validated. Furthermore, they highlight the importance of employing precise numerical techniques or other well-defined approximations for accurate cross-verification for complex QD systems. These insights are crucial for interpreting experimental observations and designing novel superconducting quantum devices based on parallel DQD architectures, which hold promise for diverse applications in quantum information processing and the exploration of exotic many-body phenomena.

\begin{acknowledgements} 
	We acknowledge discussions with M. \v{Z}onda.
	This work was supported by Grant  No. 23-05263K of the Czech Science Foundation
	and the National Science Centre (Poland) through the grant No.\ 2022/04/Y/ST3/00061.
	Computing time at the Pozna\'{n} Supercomputing and Networking Center is also acknowledged.
\end{acknowledgements}

\newpage 

\section{Appendix}

\subsection{Particle-hole transformations \label{sec:app_ph}}

The present  model possesses several symmetries when $U_1=U_2 \equiv U$ and $\Gamma_1=\Gamma_2$.
The trivial one permutes the dot indices $j$ as $1 \leftrightarrow 2$ and shall be denoted as $\mathcal{T}_0$. Because of two QDs, two independent ph transformations may be defined. The first one, does not include any relative sign between the QD's electrons and reads
\begin{align}
	\mathcal{T}_1: \hspace{1.1cm}
	d_{j\sigma}^{\vphantom{\dagger}} \rightarrow d_{j-\sigma}^{\dagger}, 
	\, \,
	c_{\mathbf{k}\sigma}^{\vphantom{\dagger}} \rightarrow c_{\mathbf{k}-\sigma}^{\dagger},
\end{align}
while adding the relative sign leads to  
\begin{align}
	\mathcal{T}_2: \hspace{0.2cm}
	d_{j\sigma}^{\vphantom{\dagger}} \rightarrow (-1)^j d_{j-\sigma}^{\dagger},
	\, \,
	c_{\mathbf{k}\sigma}^{\vphantom{\dagger}} \rightarrow c_{\mathbf{k}-\sigma}^{\dagger}.
\end{align}
Note that spin flip is incorporated into the definition of both transformations as an advantageous convention that allows both ph transformations to leave the BCS gap parameter untouched, as demonstrated below in more detail.

Applying first $\mathcal{T}_1$, Hamiltonian \eqref{eq:totalH} becomes
\begin{align}
	H
	&\rightarrow
	\left(
	-\delta_{1}-\frac{U}{2}
	\right)
	n_{1}
	+U
	n_{1\uparrow}
	n_{1\downarrow}
	\nonumber
	\\
	&\hspace{0.1cm}
	+
	\left(
	-\delta_{2}-\frac{U}{2}
	\right)
	n_{2}
	+
	U
	n_{2\uparrow}
	n_{2\downarrow}
	+
	\delta_1
	+
	\delta_2
	+
	U
	\nonumber
	\\
	&-t
	\sum_{\sigma}
	\left(
	d_{1\sigma}^{\dagger}
	d_{2\sigma}^{\vphantom{\dagger}}
	+
	\textit{h.c.}
	\right)
	-
	\sum_{j\mathbf{k}}
	V_{j,\mathbf{k}}^{\vphantom{\dagger}}
	\left(
	c^{\dagger}_{\mathbf{k}\sigma}
	d_{j\sigma}^{\vphantom{\dagger}}
	+
	\textit{h.c.}
	\right)
	\nonumber
	\\
	&-
	\sum_{\mathbf{k}\sigma}
	\varepsilon_{\mathbf{k}}
	c^{\dagger}_{\mathbf{k}\sigma}
	c_{\mathbf{k}\sigma}^{\vphantom{\dagger}}
	+
	2\sum_{\mathbf{k}}
	\varepsilon_{\mathbf{k}}
	+
	\sum_{\mathbf{k}}
	\Delta
	\left(
	c^{\dagger}_{\mathbf{k}\uparrow}
	c^{\dagger}_{-\mathbf{k}\downarrow}
	+
	\textit{h.c.}
	\right),
	\nonumber
	\\
\end{align}
with $n_{j\sigma} = d_{j\sigma}^\dag d_{j\sigma}$ and $n_j=n_{j\downarrow}+n_{j\uparrow}$,
which not only changes signs of $\delta_1$, $\delta_2, V_{1,\mathbf{k}}, V_{2,\mathbf{k}}$ and $\varepsilon_{\mathbf{k}}$ but additionally also flips the sign of the inter-dot hopping parameter $t$. Symbolically, we may summarize the action of the ph-transformation $\mathcal{T}_1$ on Hamiltonian \eqref{eq:totalH} as
\begin{align}
	\mathcal{T}_1
	&\left[
	H
	\left(
	\delta_1,\delta_2,V_{1,\mathbf{k}},V_{2,\mathbf{k}},\Delta,U,t,\varepsilon_{\mathbf{k}}
	\right)
	\right]
	\nonumber
	\\
	&=H
	\left(
	-\delta_1,-\delta_2,-V_{1,\mathbf{k}},-V_{2,\mathbf{k}},\Delta,U,-t,-\varepsilon_{\mathbf{k}}
	\right)
	\nonumber
	\\
	&+
	\left(\delta_1+\delta_2\right)
	+
	\Big(
	U + 2\sum_{\mathbf{k}}
	\varepsilon_{\mathbf{k}}
	\Big).
\end{align}
Similarly for $\mathcal{T}_2$, we obtain
\begin{align}
	\mathcal{T}_2
	&\left[
	H
	\left(
	\delta_1,\delta_2,V_{1,\mathbf{k}},V_{2,\mathbf{k}},\Delta,U,t,\varepsilon_{\mathbf{k}}
	\right)
	\right]
	\nonumber
	\\
	&=H
	\left(
	-\delta_1,-\delta_2,V_{1,\mathbf{k}},-V_{2,\mathbf{k}},\Delta,U,t,-\varepsilon_{\mathbf{k}}
	\right)
	\nonumber
	\\
	&+
	\left(\delta_1+\delta_2\right)
	+
	\Big(
	U + 2\sum_{\mathbf{k}}
	\varepsilon_{\mathbf{k}}
	\Big).
\end{align}

The extra minus sign in front of $\varepsilon_{\mathbf{k}}$ appearing in the ph-transformations $\mathcal{T}_1$ and $\mathcal{T}_2$ has no physical impact, as can be seen from the self-energy expression~\eqref{eq:Sigma_k}. However, both $\mathcal{T}_1$ and $\mathcal{T}_2$ introduce additional terms, $U+2\sum_{\mathbf{k}} \varepsilon_{\mathbf{k}}$ and $\delta_1+\delta_2$, into the transformed Hamiltonians. While the first term represents a mere constant energy shift for all parameters, which is of no significance, the latter depends on the detunings and needs a more careful treatment. Consequently, neither $\mathcal{T}_1$ nor $\mathcal{T}_2$ represents true symmetries of $H$ for general values of $\delta_1$ and $\delta_2$. The only ph-symmetric point occurs at $\delta_1=\delta_2=0$, provided that a sign change in $t$ is performed due to the transformation $\mathcal{T}_1$, or alternatively, a relative sign between $V_{1,\mathbf{k}}$ and $V_{2,\mathbf{k}}$ is introduced due to $\mathcal{T}_2$. Notably, a simultaneous sign reversal of $V_{1,\mathbf{k}}$ and $V_{2,\mathbf{k}}$ arising from $\mathcal{T}_1$ is physically irrelevant.

We emphasize that the extra energy terms generated by $\mathcal{T}_1$ or $\mathcal{T}_2$ just shift the spectrum of energy eigenstates at given $\delta_1$ and $\delta_2$ and leave the spectrum otherwise unaltered. Thus, the phase diagrams are governed by symmetries generated by $\mathcal{T}_1$ and $\mathcal{T}_2$. Explicitly, from $\mathcal{T}_1$, we conclude that simultaneous sign reversals in $\delta_1$, $\delta_2$, and $t$ leave a given phase diagram unchanged. Similarly, from $\mathcal{T}_2$, a relevant extra minus sign between $V_{1,\mathbf{k}}$ and $V_{2,\mathbf{k}}$, accompanied by sign reversals of $\delta_1$ and $\delta_2$, once again returns a symmetry operation on a given phase diagram.

Now, let us address how ph-transformations affect the AL theory. In this respect, the origin of $\Gamma_{\mathrm{DAR},j}^{\mathrm{AL}}$ and $\Gamma_{\mathrm{CAR}}^{\mathrm{AL}}$ needs thorough understanding. In the symmetric scenario considered here, they follow from the full Hamiltonian \eqref{eq:totalH} and have therefore the same magnitude and sign. However, once the $\mathcal{T}_2$ transformation is applied the new Hamiltonian acquires an extra minus sign between $V_{1,\mathbf{k}}$ and $V_{2,\mathbf{k}}$, which reverses the sign of $\Gamma_{\mathrm{\mathrm{CAR}}}^{\mathrm{AL}}$. We also emphasize that the $c$-electron transformations in $\mathcal{T}_1$ and $\mathcal{T}_2$ are no longer required in the AL theory and are strictly speaking not present at all in $\mathcal{T}_1$ and $\mathcal{T}_2$. Consequently, acting with $\mathcal{T}_1$ upon the AL Hamiltonian gives
\begin{align}
	\mathcal{T}_1
	&\left[
	H^{\mathrm{AL}}
	\left(
	\delta_1,\delta_2,\Gamma_{\mathrm{DAR}}^{\mathrm{AL}},\Gamma_{\mathrm{CAR}}^{\mathrm{AL}},U,t
	\right)
	\right]
	\nonumber
	\\
	&=H^{\mathrm{AL}}
	\left(
	-\delta_1,-\delta_2,\Gamma_{\mathrm{DAR}}^{\mathrm{AL}},\Gamma_{\mathrm{CAR}}^{\mathrm{AL}},U,-t
	\right)
	+
	\left(\delta_1+\delta_2\right).
\end{align}
Notably, both coupling strengths, $\Gamma_{\mathrm{DAR}}^{\mathrm{AL}}$ and $\Gamma_{\mathrm{CAR}}^{\mathrm{AL}}$,
retain the same relative sign. The $\mathcal{T}_2$ transformation, however, acts as
\begin{align}
	\mathcal{T}_2
	&\left[
	H^{\mathrm{AL}}
	\left(
	\delta_1,\delta_2,\Gamma_{\mathrm{DAR}}^{\mathrm{AL}},\Gamma_{\mathrm{CAR}}^{\mathrm{AL}},U,t
	\right)
	\right]
	\nonumber
	\\
	&=H^{\mathrm{AL}}
	\left(
	-\delta_1,-\delta_2,\Gamma_{\mathrm{DAR}}^{\mathrm{AL}},-\Gamma_{\mathrm{CAR}}^{\mathrm{AL}},U,t
	\right)
	+
	\left(\delta_1+\delta_2\right)
\end{align}
and introduces a relative sign between $\Gamma_{\mathrm{DAR}}^{\mathrm{AL}}$ and $\Gamma_{\mathrm{CAR}}^{\mathrm{AL}}$. As with phase diagrams based on full Hamiltonian \eqref{eq:totalH}, only the $\delta_1=\delta_2=0$ point remains truly ph-symmetric,
provided that simultaneous sign changes in $t$ and relative signs between $\Gamma_{\mathrm{DAR}}^{\mathrm{AL}}$ and $\Gamma_{\mathrm{CAR}}^{\mathrm{AL}}$ are introduced for $\mathcal{T}_1$ and $\mathcal{T}_2$, respectively. However, similarly as before, the addition of $\delta_1+\delta_2$ term does not change the relative eigenenergies of $H^{\mathrm{AL}}$ so the phase diagrams retain the symmetry observed previously for the full Hamiltonian.

The situation for ZBW approximation is essentially identical, since only $\mathbf{k}$-dependency is dropped compared to the full description. This affects only the definitions of $\mathcal{T}_1$ and $\mathcal{T}_2$, where the $\mathbf{k}$-dependency vanishes too. Nevertheless, the arguments applied to the full Hamiltonian before remain almost unaltered in ZBW. Thus, the ph-transformations $\mathcal{T}_1$ and $\mathcal{T}_2$ act upon the phase diagrams in the same way. In the case of eZBW, the additional hopping parameter needs to be taken adequately into account. 

\begin{figure*}
	\includegraphics[width=2\columnwidth]{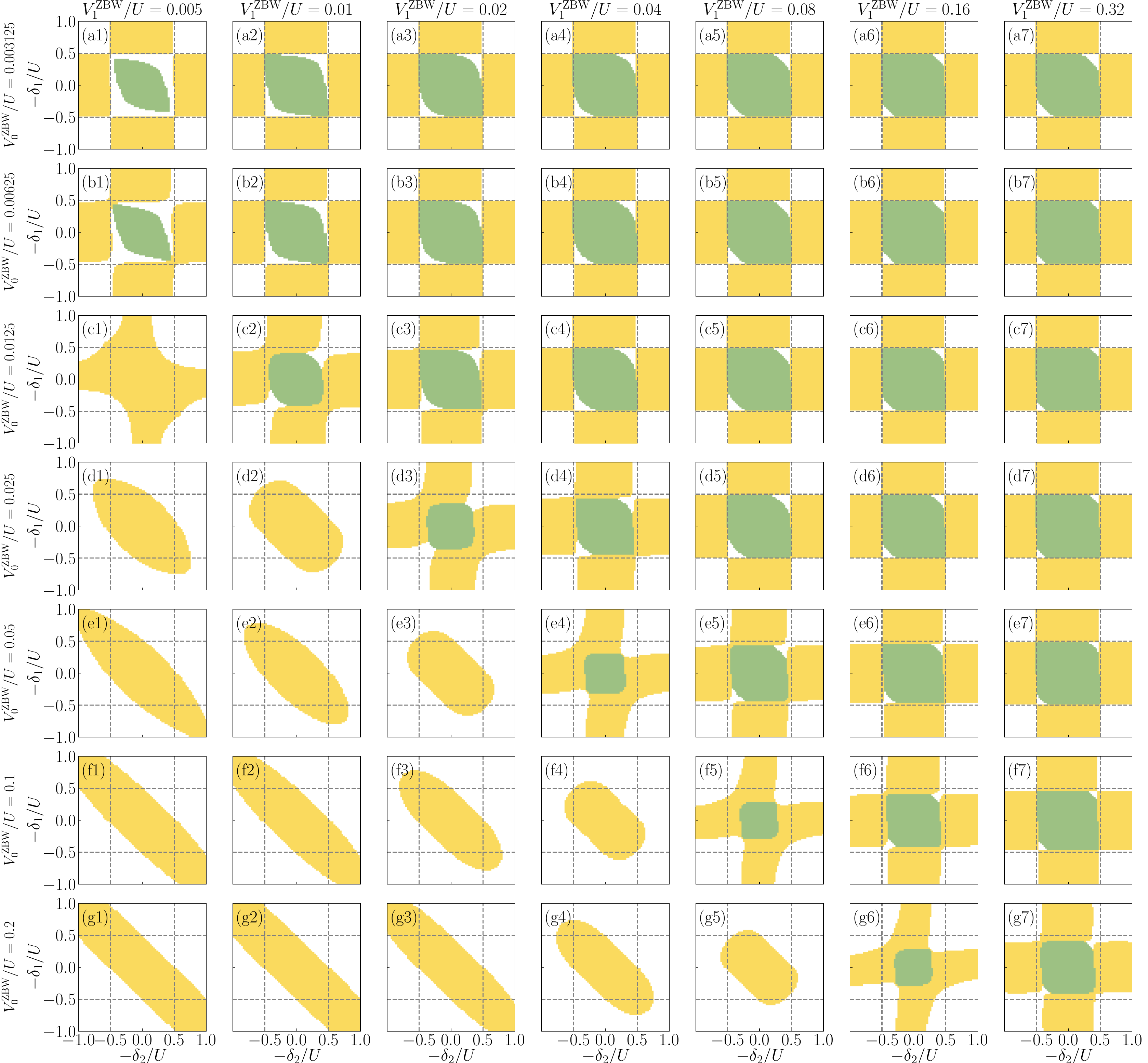}
	\caption{ 
		The series of phase diagrams calculated by eZBW in the coupling-symmetric scenario for two identical dots ($U_1=U_2=U$) with vanishing inter-dot interaction, matching the bare parameters of the exact NRG calculations in Fig.\ref{fig:nrg_UG_20}. The BCS gap parameter is set to $\Delta/U = 5 \times 10^{-2}$. Consequently, $V^{\mathrm{ZBW}}_0$ and $V^{\mathrm{ZBW}}_1$ remain as effective parameters to be freely adjusted.
		The panels are ordered horizontally by increasing values of $V^{\mathrm{ZBW}}_0$ while vertically $V^{\mathrm{ZBW}}_1$ increases from the top to the bottom. Panel (c3) matches qualitatively the exact results of Fig.~\ref{fig:nrg_UG}(d) very well. The corresponding effective values for the eZBW model are then $V^{\mathrm{ZBW}}_0/U=0.0125$ and $V^{\mathrm{ZBW}}_1/U= 0.02$.
		\label{fig:ezbw_0.05}}
\end{figure*}

\begin{figure*}
	\includegraphics[width=2\columnwidth]{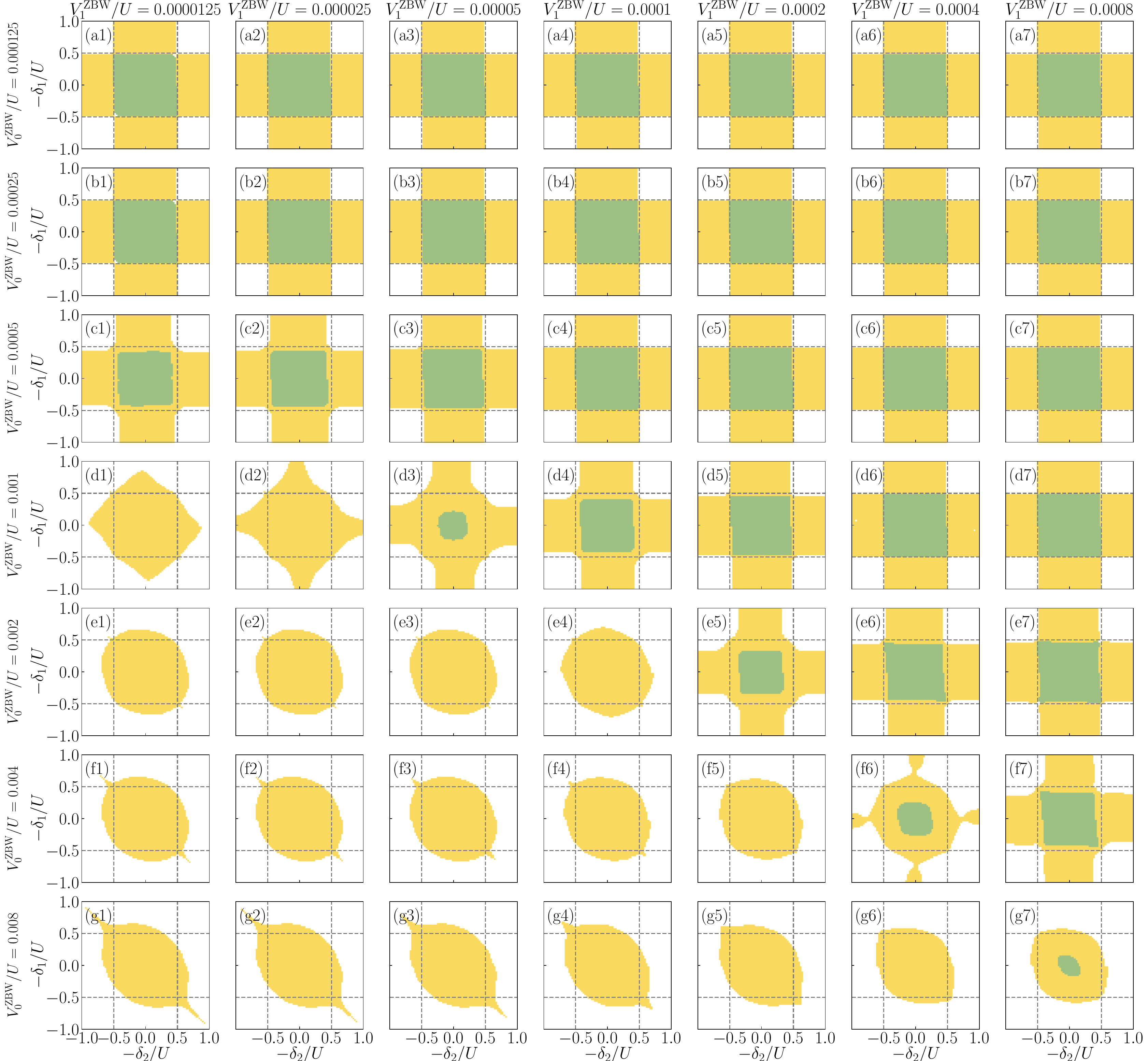}
	\caption{ 
		The series of phase diagrams calculated by eZBW in the coupling-symmetric scenario for two identical dots ($U_1=U_2=U$) with vanishing inter-dot interaction, matching the bare parameters of the exact NRG calculations in Fig.\ref{fig:nrg_UG_20}. The BCS gap parameter is set to $\Delta/U=5 \times 10^{-4}$. Consequently, $V^{\mathrm{ZBW}}_0$ and $V^{\mathrm{ZBW}}_1$ remain as effective parameters to be freely adjusted. The panels are ordered horizontally by increasing values of $V^{\mathrm{ZBW}}_0$ while vertically $V^{\mathrm{ZBW}}_1$ increases from the top to the bottom. Panel (d3) matches qualitatively the exact results of Fig.~\ref{fig:nrg_UG}(d) very well. The corresponding effective values for the eZBW model are then $V^{\mathrm{ZBW}}_0/U=0.001$ and $V^{\mathrm{ZBW}}_1/U=5 \times 10^{-5}$.
		\label{fig:ezbw_0.0005}}
\end{figure*}

\subsection{Phase diagrams in eZBW approximation \label{sec:app_nrg}}

eZBW represents a multi-parametric model with several unknown effective values. Here, we assume $V^{\mathrm{ZBW}}_1$ to be independent of the level detunings $\delta_1$ and $\delta_2$, and treat $\Delta$, $U_1$ and $U_2$ by taking their bare values as the input. Consequently, $V^{\mathrm{ZBW}}_0$, $V^{\mathrm{ZBW}}_1$, and possibly $t$ remain to be freely adjusted. We furthermore focus on coupling-symmetric scenario with $\Gamma/U=20$ and identical dots ($U_1=U_2=U$) with vanishing inter-dot interaction, matching the bare parameters of the exact NRG calculations in Fig.\ref{fig:nrg_UG_20}. Two cases of $\Delta/U = 5 \times 10^{-4}$ and $\Delta/U = 5 \times 10^{-2}$ are considered here. The parameter scan is then shown in Figs.~\ref{fig:ezbw_0.05} and \ref{fig:ezbw_0.0005} with panels ordered by increasing $V^{\mathrm{ZBW}}_0$ horizontally and $V^{\mathrm{ZBW}}_1$ vertically.

\end{document}